\documentclass{IEEEtaes}
\usepackage{color,array,amsthm}
\usepackage{graphicx}
\usepackage{amsmath}
\usepackage{amssymb}
\usepackage{algorithm}
\usepackage{algpseudocode}
\usepackage[caption=false,font=normalsize,labelfont=sf,textfont=sf]{subfig}
\usepackage{caption}
\usepackage{cite}
\usepackage{textcomp}
\usepackage{stfloats}
\usepackage{booktabs}
\usepackage{multirow}
\usepackage{color}
\usepackage{multicol}
\usepackage[table]{xcolor}
\definecolor{hl}{gray}{0.9}

\jvol{XX}
\jnum{XX}
\jmonth{XXXXX}
\paper{1234567}
\pubyear{2022}
\doiinfo{TAES.2022.Doi Number}

\newcommand{\trace}{\text{tr}}
\newcommand{\diag}{\text{diag}}
\newcommand{\transpose}{\mathsf{T}}
\newcommand{\norm}[1]{\left\lVert #1 \right\rVert}

\algrenewcommand\algorithmicrequire{\textbf{Input:}}
\algrenewcommand\algorithmicensure{\textbf{Output:}}

\makeatletter
\renewcommand{\p@subsection}{\thesection-}
\makeatother

\begin{document}

\title{Hessian-matching Based Weighting for Attitude Determination Using Short-Range DoA Measurements with IMU Assistance}
\author{Chenxin Tu}
\member{Graduate Student Member, IEEE}
\affil{Department of Electronic Engineering, Tsinghua University, Beijing, China} 

\author{Hengchuan Zhang}
\affil{Department of Electronic Engineering, Tsinghua University, Beijing, China} 

\author{Xiaowei Cui}
\affil{Department of Electronic Engineering, Tsinghua University, Beijing, China \\
State Key Laboratory of Space Network and Communication, Tsinghua University, Beijing, China
} 

\author{Gang Liu}
\affil{Department of Electronic Engineering, Tsinghua University, Beijing, China}

\author{Mingquan Lu}
\affil{Department of Electronic Engineering, Tsinghua University, Beijing, China \\
State Key Laboratory of Space Network and Communication, Tsinghua University, Beijing, China
}

\receiveddate{Manuscript received XXXXX 00, 0000; revised XXXXX 00, 0000; accepted XXXXX 00, 0000.\\
This work is supported by the National Natural Science Foundation of China under Grant No. 62541105 and the National Key R\&D Program of China under Grant No. 2021YFA0716603.}

\corresp{{\itshape (Corresponding author: Xiaowei Cui;Mingquan Lu)}}

\authoraddress{Chenxin Tu, Hengchuan Zhang and Gang Liu are with the Department of Electronic Engineering, Tsinghua University, Beijing 100084, China (e-mail: tcx22@mails.tsinghua.edu.cn; hc-zhang24@mails.tsinghua.edu.cn; liu\_gang@tsinghua.edu.cn). Xiaowei Cui and Mingquan Lu are with the Department of Electronic Engineering and the State Key Laboratory of Space Network and Communication, Tsinghua University, Beijing 100084, China (e-mail: cxw2005@tsinghua.edu.cn; lumq@tsinghua.edu.cn).}

\markboth{AUTHOR ET AL.}{SHORT ARTICLE TITLE}
\maketitle
\begin{abstract}
Accurate and reliable attitude determination (AD) is essential for unmanned vehicles operating in Global Navigation Satellite System (GNSS)-denied environments. Short-range wireless arrays can provide direction-of-arrival (DoA) measurements from multiple anchors, enabling AD by aligning corresponding direction vectors (DVs) expressed in the body and navigation frames.
In short-range scenarios, navigation-frame DVs inherit non-negligible uncertainty induced by anchor/vehicle position errors in addition to DoA-induced errors in body-frame DVs. Moreover, due to projection and unit-norm normalization, the DV errors are generally anisotropic, which motivates a total least squares (TLS) viewpoint. This paper identifies the key modeling distinction in short-range AD, develops a TLS-consistent formulation based on the total DV error and solves the resulting covariance-weighted orthogonal Procrustes problem via a manifold Gauss--Newton method. To retain the efficiency and numerical robustness of the closed-form weighted Wahba solution, we further propose Hessian-matching based scalar weighting strategies that approximate the Hessian of Wahba formulation to the TLS formulation, including a full-attitude strategy for overall accuracy and a direction-of-interest (DOI) strategy for prioritizing a selected attitude component. Finally, we incorporate IMU-derived gravity as an additional DV pair for static initialization, leading to extended Wahba and extended TLS formulations. Simulation results demonstrate that the proposed Hessian-matching weighting improves accuracy and robustness compared with existing baselines, and that gravity-DV augmentation further reduces attitude errors and improves solution availability under limited anchor availability.
\end{abstract}


\section{INTRODUCTION} \label{introduction}
Unmanned vehicles have been increasingly deployed in applications such as environmental mapping, logistics, precision agriculture, and infrastructure inspection \cite{klemas2015coastal,li2022application,velusamy2021unmanned,ellenberg2015use}. Accurate pose (position and attitude) is essential for navigation, control, and coordination of single vehicles and swarms \cite{oh2013formation,senanayake2016search}. In many missions, vehicles may be pre-positioned and required to start operating immediately (e.g., rapid deployment or coordinated takeoff), where reliable initial pose is a prerequisite for subsequent state estimation and tracking \cite{hol2009tightly,feng2020kalman}. While an initial position can often be obtained from short-range infrastructure \cite{tu2025parameterized,tu2025decoupled}, initializing attitude remains more challenging.

Attitude determination (AD) is fundamentally a vector alignment problem: it estimates the rotation between a body frame and a reference/navigation frame by aligning at least two non-collinear vectors expressed in both frames \cite{liu2022research}. Different AD sensors and systems mainly differ in how the vector pairs are constructed. For example, a star tracker provides line-of-sight unit vectors to known stars, yielding vector pairs between the camera/body frame and an inertial reference frame \cite{liebe1995star}. A magnetometer measures the local magnetic-field direction, which can be compared with the corresponding direction predicted by a geomagnetic-field model \cite{lovera2006global}. Inertial measurement units (IMUs) provide specific-force measurements that are closely tied to gravity. Since the local gravity direction is known a priori in the navigation frame, the accelerometer output can be normalized to form a gravity-direction vector observation that naturally defines a body--navigation vector pair \cite{titterton2004strapdown}. For multi-antenna Global Navigation Satellite System (GNSS) receivers, attitude can be obtained either from carrier-phase baseline vector estimation \cite{teunissen2017springer} or from satellite direction-of-arrival (DoA) processing \cite{meurer2012robust,daneshmand2014precise}. In the baseline-based approach, the relative positions (baselines) between antenna phase centers are estimated in the navigation frame and then aligned with the corresponding known baselines in the body frame determined by the antenna installation geometry. In the DoA-based approach, the antenna array estimates the incident direction vectors (DVs) in the body frame, while the corresponding reference DVs are computed from satellite--receiver geometry in the navigation frame.

Among these vector-based AD modalities, GNSS-based methods have been particularly influential for unmanned vehicles due to their widespread availability and the high accuracy achievable with multi-antenna configurations. However, when GNSS signals are blocked or severely degraded \cite{dominguez2016performance}, these GNSS-based solutions may become unavailable or unreliable, motivating complementary AD methods for GNSS-denied environments. Recent advances in short-range wireless sensing have enabled compact arrays to provide DoA (or angle-of-arrival, AoA) measurements using diverse physical layers such as ultra-wideband (UWB) and Bluetooth \cite{dotlic2017angle,li2024high,xiao2024research}. When multiple fixed anchors are available, DoA measurements yield DVs in the body frame, while anchor/vehicle position information yields corresponding DVs in the navigation frame. 

A key distinction of short-range DoA-based AD from conventional long-range counterpart (such as GNSS) is that, in short-range scenarios, the reference DVs are no longer nearly error-free: they inherit non-negligible uncertainty from anchor and vehicle position errors, in addition to the DoA-induced uncertainty in the observation DVs. As a result, the ordinary least squares (OLS) viewpoint implicitly adopted in many DoA-based formulations \cite{wang2025attitude} is generally violated, and a total least squares (TLS) viewpoint that accounts for errors in both vector sets becomes more appropriate \cite{chang2015total}. 

The canonical formulation for vector-based AD is the Wahba problem \cite{wahba1965least}, which is typically solved in a weighted form to accommodate heterogeneous measurement quality \cite{markley1988attitude}. However, the standard weighted Wahba formulation assigns a single scalar weight to each vector pair, implicitly assuming isotropic uncertainty for each vector pair \cite{chang2015total}. In practical scenarios, DV errors are often anisotropic due to projection and unit-norm normalization, as well as anisotropic DoA estimation accuracy, and their covariances can be singular (or near singular), which complicates both statistical modeling and estimator design. As a result, it remains nontrivial to select scalar weights that faithfully capture the joint effects of DoA-induced and position-induced DV uncertainties under general, geometry-dependent, and even possibly anisotropic, error statistics.

In addition to DoA-derived DVs, gravity provides a natural reference direction for attitude estimation. During static initialization, the accelerometer output can be normalized to yield a body-frame gravity-direction observation, which can be paired with the known gravity direction in the navigation frame to form an additional DV constraint \cite{titterton2004strapdown}. Most existing IMU-related approaches use this gravity-direction observation either for a standalone leveling step \cite{ahmed2017accurate} or, together with another reference direction (e.g., magnetic field or velocity), to deterministically recover attitude \cite{decelis2018attitude,gebre2000gyro}, rather than treating gravity as a homogeneous DV pair that can be jointly aligned with DoA-derived DVs within the same solver. Moreover, a principled yet computationally efficient scheme for weighting the gravity DV relative to the DoA-derived DV pairs remains insufficiently explored.

This paper addresses short-range DoA-based AD by developing a TLS-consistent formulation based on the \emph{total DV error} and by designing Hessian-matching based scalar weights for closed-form Wahba solutions. In addition, for static initialization, we incorporate IMU-derived gravity information as an additional DV pair in the same DV-alignment framework to improve solution availability and robustness. The main contributions are summarized as follows:
\begin{itemize}
    \item \emph{TLS-consistent modeling with anisotropic DV errors:} We formulate short-range DoA-based AD by explicitly modeling the total DV error induced jointly by DoA estimation and position-derived DV uncertainties, allowing general (anisotropic and geometry-dependent) covariance structures. The resulting covariance-weighted orthogonal Procrustes problem is solved via a manifold Gauss--Newton method on $\mathrm{SO}(3)$.
    \item \emph{Hessian-matching based scalar weighting for closed-form Wahba:} To avoid the computational burden and potential sensitivity of iterative TLS solvers, we design practical scalar weights for the closed-form weighted Wahba solver by matching the local Hessian geometry implied by the TLS model. We develop (i) a full-attitude strategy that targets overall accuracy and (ii) a direction-of-interest (DOI) strategy that prioritizes a user-selected attitude component.
    \item \emph{IMU-assisted gravity-DV augmentation with analytic weighting:} We incorporate IMU-derived gravity as an additional DV pair within the same DV-alignment solver, leading to extended Wahba (EWahba) and extended TLS (ETLS) formulations. The gravity weight follows the same Hessian-matching principle and can be determined analytically from accelerometer error statistics.
\end{itemize}

The remainder of this paper is organized as follows. Section~\ref{preliminaries} establishes the theoretical foundations by reviewing the Wahba problem and gravity DV estimation. Section~\ref{problem-formulation} introduces the problem setup and formulates the objectives. Section~\ref{method1} presents the TLS-consistent formulation and Hessian-matching based weighting strategies for short-range DoA-based AD. Section~\ref{method2} develops the IMU-assisted methods via gravity-DV augmentation. Section~\ref{simulation} reports and discusses simulation results, and Section~\ref{conclusion} concludes the paper.

Notation: Vectors are denoted by bold lowercase letters, and matrices by bold uppercase letters. The transpose and inverse of a matrix are denoted by $(\cdot)^{\transpose}$ and $(\cdot)^{-1}$, respectively. The trace and determinant of a matrix are indicated by $\trace(\cdot)$ and $\det(\cdot)$, respectively. The operator $\diag(\cdot)$ constructs a diagonal matrix from its input vector. $[\cdot]_{\times}$ denotes the skew-symmetric matrix corresponding to the inner vector, which is associated with the cross product operation. The Euclidean norm is represented by $\norm{\cdot}$, and $\boldsymbol{I}$ denotes the identity matrix of appropriate dimensions. The expectation operator is denoted by $\mathbb{E}(\cdot)$. The special orthogonal group is denoted by $\mathrm{SO}(n)=\{\boldsymbol{R}\in\mathbb{R}^{n\times n}\mid \boldsymbol{R}^{\transpose}\boldsymbol{R}=\boldsymbol{I},\ \det(\boldsymbol{R})=1\}$. Unless otherwise specified, we use $\hat{({\cdot})}$ to denote estimated (or predicted) quantities, $\tilde{({\cdot})}$ to denote measured quantities and $\breve{({\cdot})}$ to denote regularized quantities.

\section{PRELIMINARIES} \label{preliminaries}

In this section, we introduce the essential preliminaries required for the subsequent development, including the Wahba problem and gravity DV estimation derived by an IMU during static initialization. These concepts provide the theoretical foundation for the proposed IMU-assisted DoA-based AD framework.

\subsection{The Wahba Problem}\label{sec:wahba}
In this subsection, we briefly review the Wahba problem and its weighted formulation, which form the mathematical basis of vector-based AD. Consider two sets of $K$ corresponding 3-D unit vectors expressed in an observation frame and a reference frame, denoted by $\{\boldsymbol{o}_{k}\}_{k=1}^{K}$ and $\{\boldsymbol{r}_{k}\}_{k=1}^{K}$, respectively, and referred to as the \emph{observation vectors} and \emph{reference vectors}. Their physical meanings and constructions in the short-range scenarios will be specified in Section~\ref{problem-formulation}. The attitude is represented by a rotation matrix $\boldsymbol{R}\in \mathrm{SO}(3)$ that transforms vectors from the reference frame to the observation frame. In the ideal noise-free case, the two vector sets satisfy $\boldsymbol{o}_{k}=\boldsymbol{R}\boldsymbol{r}_{k}$ for $k=1,\ldots,K$. In practice, however, only noisy measurements $\tilde{\boldsymbol{o}}_{k}$ and $\tilde{\boldsymbol{r}}_{k}$ are available. The Wahba problem estimates $\boldsymbol{R}$ by aligning the two vector sets in a least-squares sense, and is formulated as \cite{wahba1965least}
\begin{equation}
\hat{\boldsymbol{R}}
= \arg \min_{\boldsymbol{R}\in \mathrm{SO}(3)}
\sum_{k=1}^{K}\norm{\tilde{\boldsymbol{o}}_{k}-\boldsymbol{R}\tilde{\boldsymbol{r}}_{k}}^2.
\label{eq1}
\end{equation}
When the measurement qualities are heterogeneous across $k$, a weighted form is typically adopted \cite{markley1988attitude}:
\begin{equation}
\hat{\boldsymbol{R}}
= \arg \min_{\boldsymbol{R}\in \mathrm{SO}(3)}
\sum_{k=1}^{K} w_k \norm{\tilde{\boldsymbol{o}}_{k}-\boldsymbol{R}\tilde{\boldsymbol{r}}_{k}}^2,
\label{eq2}
\end{equation}
where $w_k\ge 0$ denotes the weight of the $k$-th vector pair.

The weighted Wahba problem admits an efficient closed-form solution based on singular value decomposition (SVD) \cite{markley1988attitude}. The solution procedure is summarized as follows:
\begin{enumerate}
    \item Construct the $3\times 3$ matrix
    \begin{equation}
        \boldsymbol{B} = \sum_{k=1}^{K} w_k \tilde{\boldsymbol{o}}_k (\tilde{\boldsymbol{r}}_k)^{\transpose}.
        \label{eq3}
    \end{equation}
    \item Compute the SVD of $\boldsymbol{B}$:
    \begin{equation}
        \boldsymbol{B} = \boldsymbol{U}\boldsymbol{S}\boldsymbol{V}^{\transpose}.
        \label{eq4}
    \end{equation}
    \item Obtain the rotation estimate as
    \begin{equation}
        \hat{\boldsymbol{R}} = \boldsymbol{U}\boldsymbol{M}\boldsymbol{V}^{\transpose},
        \label{eq5}
    \end{equation}
    where $\boldsymbol{M}$ enforces $\hat{\boldsymbol{R}}\in \mathrm{SO}(3)$:
    \begin{equation}
        \boldsymbol{M} = \diag \left( \left[ 1,1,\det(\boldsymbol{U}) \det(\boldsymbol{V}) \right] \right).
        \label{eq6}
    \end{equation}
\end{enumerate}
In addition to the attitude estimate itself, the attitude error covariance can also be derived \cite{markley1988attitude}. This covariance information is of significant importance for performance prediction and for initializing filter-based state estimation and tracking algorithms \cite{hol2009tightly, feng2020kalman}. The detailed derivation and results can be found in \cite{markley1988attitude} and are therefore omitted here for brevity.

\subsection{Gravity DV Estimation} \label{sec:leveling}
Let $\tilde{\boldsymbol{f}}^{b}=[\tilde f_x,\tilde f_y,\tilde f_z]^{\transpose}$ denote the temporal-averaged accelerometer output expressed in the body frame. During static initialization, the measured specific force is dominated by gravity; in the ideal noise-free case, $\boldsymbol{f}^{b} = -\boldsymbol{g}^{b}$, where $\boldsymbol{g}^{b}$ denotes the gravity vector in the body frame. The gravity vector in the navigation frame is $\boldsymbol{g}^{n}=[0,0,-g]^{\transpose}$\footnote{Throughout this paper, the navigation frame is defined as the East--North--Up (ENU) frame, and the body frame follows the Right--Front--Up (RFU) convention.}. Therefore, an IMU-derived unit gravity DV pair can be constructed as
\begin{equation}
    \tilde{\boldsymbol{v}}_{\boldsymbol{g}}^{b} \triangleq \frac{-\tilde{\boldsymbol{f}}^{b}}{\norm{\tilde{\boldsymbol{f}}^{b}}},
    \quad
    \boldsymbol{v}_{\boldsymbol{g}}^{n} \triangleq \frac{\boldsymbol{g}^{n}}{g}=[0,0,-1]^{\transpose},
    \label{eq7}
\end{equation}
which will be incorporated as an additional vector pair in the DV-alignment framework for IMU-assisted attitude estimation in Section~\ref{method2}.

\section{Problem Formulation} \label{problem-formulation}
\subsection{Problem Setup}
\begin{figure}
\centering
\includegraphics[width=\linewidth]{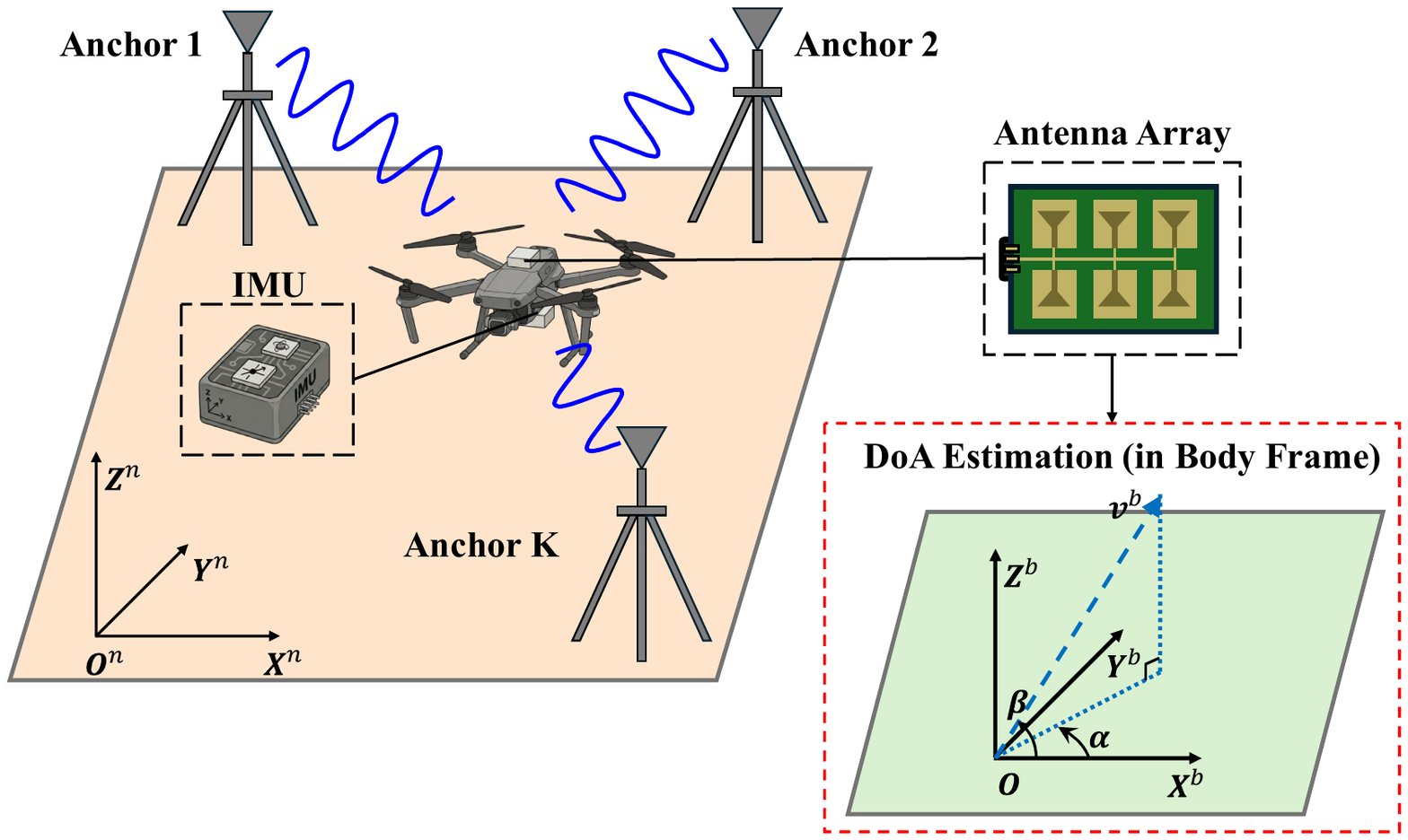}
\caption{ Problem setup for short-range DoA-based AD of an unmanned vehicle. An onboard DoA-capable array receives signals from $K$ fixed anchors. The vehicle attitude is recovered by aligning the body-frame DoA-derived DVs with the navigation-frame position-derived DVs, with optional assistance from an onboard IMU.}
\label{fig2}
\end{figure}

As illustrated in Fig.~\ref{fig2}, we consider an unmanned vehicle equipped with a DoA-capable short-range wireless antenna array and an IMU. There are $K$ fixed anchors transmitting wireless signals that are received by the onboard array. For each anchor, the array outputs a DoA measurement parameterized by azimuth and elevation angles, denoted as $\tilde{\boldsymbol{\gamma}}_k=[\tilde{\alpha}_k,\tilde{\beta}_k]^{\transpose}$ $(k=1,\ldots,K)$. The AD task is to estimate the vehicle's attitude, represented by the rotation matrix $\boldsymbol{R}\in \mathrm{SO}(3)$ from the body frame $\mathcal{B}$ to the navigation frame $\mathcal{N}$
\footnote{In this paper, we adopt the $\mathcal{B}\to\mathcal{N}$ mapping and treat navigation-frame DVs as observation vectors and body-frame DVs as reference vectors. This convention better aligns with the misalignment-vector analysis and the Hessian-matching weighting strategies developed in Section~\ref{method1}.}.
Given the DoA measurement $\tilde{\boldsymbol{\gamma}}_k$, the corresponding unit DV expressed in the body frame $\mathcal{B}$ is constructed via the standard spherical-to-Cartesian mapping
\begin{equation}
    \tilde{\boldsymbol{v}}_k^{b} = \begin{bmatrix}
        \cos{\tilde{\beta}_k} \cos{\tilde{\alpha}_k}  \\
        \cos{\tilde{\beta}_k} \sin{\tilde{\alpha}_k}  \\
        \sin{\tilde{\beta}_k}
    \end{bmatrix}.
\label{eq8}
\end{equation}
The set $\{\tilde{\boldsymbol{v}}_k^{b}\}_{k=1}^K$ constitutes the reference vectors in the sense of Section~\ref{sec:wahba}. A concrete realization of three-element array and its error characterization are provided in \hyperref[appendix]{Appendix}.

Let $\boldsymbol{p}_k^n$ and $\boldsymbol{p}_0^n$ denote the true positions of the $k$-th anchor and the vehicle in the navigation frame $\mathcal{N}$, with estimates $\hat{\boldsymbol{p}}_{k}^n$ and $\hat{\boldsymbol{p}}_{0}^n$, respectively. The true unit DV from the vehicle to anchor $k$ in frame $\mathcal{N}$ is
\begin{equation}
    \boldsymbol{v}_{k}^{n} = \frac{ \boldsymbol{p}_{k}^{n} - \boldsymbol{p}_{0}^{n}}
    {\norm{\boldsymbol{p}_{k}^{n} - \boldsymbol{p}_{0}^{n}}},
    \label{eq9}
\end{equation}
and the corresponding measured unit DV is obtained by replacing the true positions with their estimated counterparts in \eqref{eq9}, denoted as $\tilde{\boldsymbol{v}}_k^{n}$. The set $\{\tilde{\boldsymbol{v}}_k^{n}\}_{k=1}^K$ constitutes the observation vectors in the sense of Section~\ref{sec:wahba}. Throughout this paper, the position errors $\Delta \boldsymbol{p}_m=\hat{\boldsymbol{p}}_m^n-\boldsymbol{p}_m^n$ are modeled as independent zero-mean Gaussian random vectors with covariance matrices $\boldsymbol{\Sigma}_{\boldsymbol{p}_m}=\sigma_{p_m}^2\boldsymbol{I}$ for $m=0,1,\ldots,K$.

In addition, during static initialization, an onboard IMU can provide gravity-direction information that complements the DoA-derived DVs. The accuracy of this gravity-direction observation is primarily determined by the accelerometer error characteristics. We assume that high-frequency random noise has been effectively suppressed by temporal averaging and that deterministic biases have been properly calibrated. Under these assumptions, the remaining accelerometer error is dominated by the turn-on bias, which is modeled as a zero-mean Gaussian random vector $\boldsymbol{b}_a=[b_{a_x},b_{a_y},b_{a_z}]^{\transpose}$ with covariance matrix $\boldsymbol{\Sigma}_{\boldsymbol{b}_a}=\sigma_{b_a}^2 \boldsymbol{I}$.

\subsection{Research Target}
In the noise-free case, the corresponding true vectors satisfy the rotation relation
\begin{equation}
     {\boldsymbol{v}}_{k}^{n} = {\boldsymbol{R}}{\boldsymbol{v}}_{k}^{b}, \quad k=1,\ldots,K,
 \label{eq10}
\end{equation}
Our goal is to estimate the attitude $\boldsymbol{R}\in\mathrm{SO}(3)$ (mapping $\mathcal{B}\to\mathcal{N}$) from the available measurements. Specifically, the inputs include:
\begin{itemize}
    \item the DoA measurements $\{\tilde{\boldsymbol{\gamma}}_k\}_{k=1}^{K}$, from which we construct the body-frame DVs $\{\tilde{\boldsymbol{v}}_k^{b}\}_{k=1}^{K}$ via \eqref{eq8},
    \item the anchor/vehicle position estimates $\{\hat{\boldsymbol{p}}_k^{n},\hat{\boldsymbol{p}}_0^{n}\}$, from which we construct the navigation-frame DVs $\{\tilde{\boldsymbol{v}}_k^{n}\}_{k=1}^{K}$ via \eqref{eq9}, and
    \item optionally, the accelerometer output $\tilde{\boldsymbol{f}}^{b}$, from which we construct the gravity DV pair $\left(\boldsymbol{v}_{\boldsymbol{g}}^{n},\tilde{\boldsymbol{v}}_{\boldsymbol{g}}^{b}\right)$ via \eqref{eq7}.
\end{itemize}
The output of the AD problem is the attitude estimate $\hat{\boldsymbol{R}}$.

In practice, due to measurement errors, the available DV estimates deviate from their true values. In short-range DoA-based AD, both the body-frame DVs $\{\tilde{\boldsymbol{v}}_k^{b}\}_{k=1}^{K}$ and the navigation-frame DVs $\{\tilde{\boldsymbol{v}}_k^{n}\}_{k=1}^{K}$ are subject to non-negligible errors, and their covariances can be characterized from DoA estimation and position-error statistics. Moreover, because DV construction involves projection and unit-norm normalization, the resulting DV errors are generally anisotropic and geometry-dependent. These features motivate a TLS viewpoint for modeling and estimator design. Section~\ref{method1} specifies the a covariance-weighted TLS formulation based on the total DV errors and derives Hessian-matching based scalar weights for efficient closed-form Wahba solutions. Section~\ref{method2} further incorporates the IMU-derived gravity DV as an additional vector pair to enable effective IMU-assisted AD.

\section{Short-Range DoA-based AD} \label{method1}
This section focuses on attitude estimation using only short-range DoA measurements. We first characterize the navigation-frame DV uncertainty induced by anchor/vehicle position errors. We then formulate a TLS-consistent covariance-weighted least-squares objective based on the total DV error covariance and solve it through manifold Gauss-Newton. Finally, we derive practical \emph{scalar} weights for the closed-form Wahba solver by matching the local second-order Hessian geometry implied by the TLS model, yielding a Hessian-matching based weighting rule that explicitly reflects both DoA-induced and position-induced uncertainties.

\subsection{Navigation-Frame DV Errors Induced by Position Uncertainties}
In short-range scenarios, the navigation-frame DVs $\tilde{\boldsymbol{v}}_k^{n}$ are computed from estimated anchor and vehicle positions via \eqref{eq9}. Consequently, position errors directly translate into navigation-frame DV errors. The relative position error is
\begin{equation}
    \Delta \boldsymbol{p}_{k;0} \triangleq (\hat{\boldsymbol{p}}_k^n-\hat{\boldsymbol{p}}_0^n)-(\boldsymbol{p}_k^n-\boldsymbol{p}_0^n)=\Delta \boldsymbol{p}_k-\Delta \boldsymbol{p}_0,
    \label{eq11}
\end{equation}
Denote the true anchor--vehicle distance by $d_{k;0}\triangleq\|\boldsymbol{p}_k^n-\boldsymbol{p}_0^n\|$.
By neglecting second-order error terms, the estimated navigation-frame DV admits the first-order approximation
\begin{equation}
    \tilde{\boldsymbol{v}}_k^{n} \approx 
    {\boldsymbol{v}}_{k}^{n} + \frac{\Delta \boldsymbol{p}_{k;0} - {\boldsymbol{v}}_{k}^{n}{{\boldsymbol{v}}_{k}^{n}}^{\transpose} \Delta \boldsymbol{p}_{k;0}}{d_{k;0}}
    = {\boldsymbol{v}}_{k}^{n} + \frac{\Delta \boldsymbol{p}_{k;0}^{\perp}}{d_{k;0}},
    \label{eq12}
\end{equation}
where $\Delta \boldsymbol{p}_{k;0}^{\perp} = (\boldsymbol{I}-{\boldsymbol{v}}_{k}^{n}{{\boldsymbol{v}}_{k}^{n}}^{\transpose})\Delta \boldsymbol{p}_{k;0}$ is the component of $\Delta \boldsymbol{p}_{k;0}$ orthogonal to $\boldsymbol{v}_k^n$.
Define the projection matrix onto the plane orthogonal to $\boldsymbol{v}_k^n$ as
$\boldsymbol{P}_{\boldsymbol{v}_{k}^{n}}^{\perp} \triangleq \boldsymbol{I}-{\boldsymbol{v}}_{k}^{n}{{\boldsymbol{v}}_{k}^{n}}^{\transpose}$.
Then the navigation-frame DV error is
\begin{equation}
    \Delta {\boldsymbol{v}}_{k}^{n} \triangleq \tilde{\boldsymbol{v}}_k^{n} - {\boldsymbol{v}}_{k}^{n}
    = \frac{ \boldsymbol{P}_{\boldsymbol{v}_{k}^{n}}^{\perp} \Delta \boldsymbol{p}_{k;0} }{d_{k;0}}.
    \label{eq13}
\end{equation}
With the isotropic position-error model in Section~\ref{problem-formulation}, the covariance of $\Delta\boldsymbol{v}_k^n$ is
\begin{equation}
\begin{split}
    \boldsymbol{\Sigma}_{\boldsymbol{v}_k^n} &= \mathbb{E} \left[ \Delta {\boldsymbol{v}}_{k}^{n}\left(\Delta {\boldsymbol{v}}_{k}^{n}\right)^{\transpose} \right] \\
    &= \frac{ \boldsymbol{P}_{\boldsymbol{v}_{k}^{n}}^{\perp}( \boldsymbol{\Sigma}_{\boldsymbol{p}_k}+\boldsymbol{\Sigma}_{\boldsymbol{p}_0}) \left(\boldsymbol{P}_{\boldsymbol{v}_{k}^{n}}^{\perp}\right)^{\transpose} }{d_{k;0}^2} \\
    &= \frac{(\sigma_{p_k}^2+\sigma_{p_0}^2)}{d_{k;0}^2} \boldsymbol{P}_{\boldsymbol{v}_{k}^{n}}^{\perp}.
\end{split}
 \label{eq14}
\end{equation}
Equation \eqref{eq14} shows that navigation-frame DV uncertainty increases rapidly as $d_{k;0}$ decreases, which is the key reason that the OLS assumption in the long-range case becomes invalid in short-range settings.

\subsection{TLS-Consistent Covariance-Weighted Formulation} \label{sec:tls_formulation}
The true DV vectors satisfy the rotation relation \eqref{eq10}. In practice, however, both $\tilde{\boldsymbol{v}}_k^n$ and $\tilde{\boldsymbol{v}}_k^b$ are noisy. To explicitly reflect the joint uncertainty in both vector sets, we introduce the total DV error for a candidate attitude $\widehat{\boldsymbol{R}}\in\mathrm{SO}(3)$ as
\begin{equation}
    \Delta \boldsymbol{v}_k(\widehat{\boldsymbol{R}}) \triangleq \tilde{\boldsymbol{v}}_k^{n}-\widehat{\boldsymbol{R}}\tilde{\boldsymbol{v}}_k^{b}.
    \label{eq15}
\end{equation}
Let $\Delta \boldsymbol{v}_k^n=\tilde{\boldsymbol{v}}_k^n-\boldsymbol{v}_k^n$ and $\Delta \boldsymbol{v}_k^b=\tilde{\boldsymbol{v}}_k^b-\boldsymbol{v}_k^b$ denote the observation- and reference-vector errors. Using these definitions and the noise-free relation \eqref{eq10}, \eqref{eq15} can be rewritten as the linear constraint
\begin{equation}
    \Delta \boldsymbol{v}_k(\widehat{\boldsymbol{R}}) = \Delta {\boldsymbol{v}}_{k}^{n} - \widehat{\boldsymbol{R}}\Delta {\boldsymbol{v}}_{k}^{b}.
    \label{eq16}
\end{equation}
Assuming $\Delta {\boldsymbol{v}}_{k}^{n}$ and $\Delta {\boldsymbol{v}}_{k}^{b}$ are independent zero-mean random vectors with covariance matrices $\boldsymbol{\Sigma}_{\boldsymbol{v}_k^n}$ and $\boldsymbol{\Sigma}_{\boldsymbol{v}_k^b}$, the covariance of the total DV error becomes
\begin{equation}
    \boldsymbol{\Sigma}_{\boldsymbol{v}_k}
    = \mathbb{E} \left[ \Delta \boldsymbol{v}_k \Delta \boldsymbol{v}_k^{\transpose} \right]
    = \boldsymbol{\Sigma}_{\boldsymbol{v}_k^n} + \widehat{\boldsymbol{R}}\boldsymbol{\Sigma}_{\boldsymbol{v}_k^b}\widehat{\boldsymbol{R}}^{\transpose}
    \label{eq17}
\end{equation}
where $\boldsymbol{\Sigma}_{\boldsymbol{v}_k^n}$ is given by \eqref{eq14} and $\boldsymbol{\Sigma}_{\boldsymbol{v}_k^b}$ depends on the array configuration and DoA estimation accuracy (see \hyperref[appendix]{Appendix} for an example).

To explicitly account for the fact that both vector sets are contaminated by noise, we adopt a TLS viewpoint and consider the following quadratic cost:
\begin{equation}
J_{\mathrm{TLS}}= \sum_{k=1}^K
(\Delta{\boldsymbol{v}}_{k}^{b})^{\transpose}\breve{\boldsymbol{\Sigma}}_{\boldsymbol{v}_k^b}^{-1}(\Delta {\boldsymbol{v}}_{k}^{b})
    + (\Delta {\boldsymbol{v}}_{k}^{n})^{\transpose}\breve{\boldsymbol{\Sigma}}_{\boldsymbol{v}_k^n}^{-1}(\Delta {\boldsymbol{v}}_{k}^{n})
\label{eq18}
\end{equation}
where $\breve{\boldsymbol{\Sigma}}_{\boldsymbol{v}_k^b}\triangleq \boldsymbol{\Sigma}_{\boldsymbol{v}_k^b}+\epsilon \boldsymbol{I}$ and $\breve{\boldsymbol{\Sigma}}_{\boldsymbol{v}_k^n}\triangleq \boldsymbol{\Sigma}_{\boldsymbol{v}_k^n}+\epsilon \boldsymbol{I}$ are regularized covariances with a small parameter $\epsilon>0$ introduced to ensure numerical invertibility in the presence of the unit-norm constraint. By applying the Lagrange multiplier method \cite{nocedal2006numerical} to eliminate $(\Delta {\boldsymbol{v}}_{k}^{b},\,\Delta {\boldsymbol{v}}_{k}^{n})$ under the constraint \eqref{eq16}, we obtain a reduced cost that depends only on $\widehat{\boldsymbol{R}}$ and the total DV error $\Delta \boldsymbol{v}_k$:
\begin{equation}
    \begin{split}
        J_{\mathrm{TLS}}(\widehat{\boldsymbol{R}}) = \sum_{k=1}^K
    (\tilde{\boldsymbol{v}}_k^{n}-\widehat{\boldsymbol{R}}\tilde{\boldsymbol{v}}_k^{b})^{\transpose}
        \breve{\boldsymbol{\Sigma}}_{\boldsymbol{v}_k}^{-1}
    (\tilde{\boldsymbol{v}}_k^{n}-\widehat{\boldsymbol{R}}\tilde{\boldsymbol{v}}_k^{b})
    \end{split}
\label{eq19}
\end{equation}
where $\breve{\boldsymbol{\Sigma}}_{\boldsymbol{v}_k}=\breve{\boldsymbol{\Sigma}}_{\boldsymbol{v}_k^n} + \widehat{\boldsymbol{R}}\breve{\boldsymbol{\Sigma}}_{\boldsymbol{v}_k^b}\widehat{\boldsymbol{R}}^{\transpose}$.

Minimizing \eqref{eq19} over $\widehat{\boldsymbol{R}}\in\mathrm{SO}(3)$ yields a covariance-weighted orthogonal Procrustes problem (WOPP). If the DV noise is isotropic, as commonly assumed in the literature, then the WOPP reduces to a weighted Wahba formulation \cite{chang2015total}. However, in practical scenarios, due to the unit-norm constraint and the nonlinear propagation of DoA estimation errors, DV noise is typically anisotropic and therefore not equivalent to the weighted Wahba formulation. While \eqref{eq19} can be solved iteratively by optimizing $\widehat{\boldsymbol{R}}$ on $\mathrm{SO}(3)$, such direct manifold optimization is often computationally demanding and may be sensitive to initialization. To address these challenges, we adopt a manifold Gauss--Newton approach on $\mathrm{SO}(3)$ using a Lie-algebra perturbation, which iteratively solves for a local increment in the tangent space at the current estimate and then map it back to $\mathrm{SO}(3)$ via the exponential map.

We perturb the attitude using a first-order Phi-angle misalignment model \cite{niu2021wheel}. Let $[\cdot]_{\times}$ denote the skew-symmetric (cross-product) matrix such that $[\boldsymbol{a}]_{\times}\boldsymbol{b}=\boldsymbol{a}\times\boldsymbol{b}$. Using a left-multiplicative perturbation, we write
\begin{equation}
\tilde{\boldsymbol{v}}_k^{n}-\Delta\boldsymbol{v}_k^{n}
\approx\ (\boldsymbol{I}+[\boldsymbol{\phi}]_{\times}) \, \widehat{\boldsymbol{R}} (\tilde{\boldsymbol{v}}_k^{b}-\Delta\boldsymbol{v}_k^{b}).
\label{eq20}
\end{equation}
where 
\begin{equation}
\widehat{\boldsymbol{R}}\ \approx\ (\boldsymbol{I}-[\boldsymbol{\phi}]_{\times})\,\boldsymbol{R},
\label{eq21}
\end{equation}
and $\boldsymbol{\phi}\triangleq[\phi_E,\ \phi_N,\ \phi_U]^{\transpose}\in\mathbb{R}^3$ is the small misalignment vector expressed in the navigation frame (ENU). Its components represent small rotations about the east, north, and up axes, respectively. We retain only first-order terms in $(\boldsymbol{\phi},\Delta\boldsymbol{v}_k^{n},\Delta\boldsymbol{v}_k^{b})$. Then, \eqref{eq20} becomes
\begin{equation}
\tilde{\boldsymbol{v}}_k^{n}-\Delta\boldsymbol{v}_k^{n}
\approx\ \widehat{\boldsymbol{R}}\,\tilde{\boldsymbol{v}}_k^{b}
-\widehat{\boldsymbol{R}}\,\Delta\boldsymbol{v}_k^{b}
+[\boldsymbol{\phi}]_{\times}\,\widehat{\boldsymbol{R}}\,\tilde{\boldsymbol{v}}_k^{b}.
\label{eq22}
\end{equation}
Define the predicted vector $\hat{\boldsymbol{v}}_k^{n}\triangleq\widehat{\boldsymbol{R}}\,\tilde{\boldsymbol{v}}_k^{b}$ and the difference residual
\begin{equation}
\boldsymbol{z}_k\ \triangleq\ \tilde{\boldsymbol{v}}_k^{n}-\hat{\boldsymbol{v}}_k^{n}.
\label{eq23}
\end{equation}
Using the identity $[\boldsymbol{\phi}]_{\times}\,\hat{\boldsymbol{v}}=-[\hat{\boldsymbol{v}}]_{\times}\,\boldsymbol{\phi}$, we obtain the linearized residual equation
\begin{equation}
\boldsymbol{z}_k
=  \boldsymbol{J}_k \boldsymbol{\phi}
+\Delta \boldsymbol{v}_k.
\label{eq24}
\end{equation}
where the Jacobian $\boldsymbol{J}_k$ is given by
\begin{equation}
    \boldsymbol{J}_k = -[\hat{\boldsymbol{v}}_k^{n}]_{\times}.
    \label{eq25}
\end{equation}
and $\Delta \boldsymbol{v}_k$ is the total DV error defined in \eqref{eq16} with covariance $\boldsymbol{\Sigma}_{\boldsymbol{v}_k}$ in \eqref{eq17}. Then, a covariance-weighted least-squares formulation can be constructed based on \eqref{eq24} as:
\begin{equation}
J_{\mathrm{TLS}}(\boldsymbol{\phi})
=\frac{1}{2}\sum_{k=1}^{K} (\boldsymbol{z}_k - \boldsymbol{J}_k \boldsymbol{\phi})^{\transpose}\,(\breve{\boldsymbol{\Sigma}}_{\boldsymbol{v}_k})^{-1}\,(\boldsymbol{z}_k - \boldsymbol{J}_k \boldsymbol{\phi}).
\label{eq26}
\end{equation}
At iteration $i$ with current estimate $\widehat{\boldsymbol{R}}_i$, compute
\begin{equation}
    \begin{gathered}
        \hat{\boldsymbol{v}}_{k,i}^{n} =\widehat{\boldsymbol{R}}_i\tilde{\boldsymbol{v}}_k^{b}, 
        \\
        \boldsymbol{z}_{k,i} =\tilde{\boldsymbol{v}}_k^{n}-\hat{\boldsymbol{v}}_{k,i}^{n}, 
        \\
        \boldsymbol{J}_{k,i} =-[\hat{\boldsymbol{v}}_{k,i}^{n}]_{\times}, 
        \\
        \breve{\boldsymbol{\Sigma}}_{\boldsymbol{v}_{k,i}} =\breve{\boldsymbol{\Sigma}}_{\boldsymbol{v}_k^n}+\widehat{\boldsymbol{R}}_i\breve{\boldsymbol{\Sigma}}_{\boldsymbol{v}_k^b}\widehat{\boldsymbol{R}}_i^{\transpose}, 
        \\ 
        \boldsymbol{W}_{k,i} =(\breve{\boldsymbol{\Sigma}}_{\boldsymbol{v}_k,i})^{-1}.
    \end{gathered}
    \label{eq27}
\end{equation}
Then a normal equation can be formed as
\begin{equation}
\boldsymbol{F}_{\mathrm{sys},i}\,\delta\boldsymbol{\phi}_i
=\boldsymbol{g}_{\mathrm{sys},i}
\label{eq28}
\end{equation}
where
\begin{equation}
    \begin{aligned}
\boldsymbol{F}_{\mathrm{sys},i} &= \sum_{k=1}^{K} \boldsymbol{J}_{k,i}^{\transpose}\,\boldsymbol{W}_{k,i}\,\boldsymbol{J}_{k,i}, \\
\boldsymbol{g}_{\mathrm{sys},i} &= \sum_{k=1}^{K} \boldsymbol{J}_{k,i}^{\transpose}\,\boldsymbol{W}_{k,i}\,\boldsymbol{z}_{k,i}.
    \end{aligned}
\label{eq29}
\end{equation}
and the increment $\delta\boldsymbol{\phi}_i$ can be solved as
\begin{equation}
\delta\boldsymbol{\phi}_i=\boldsymbol{F}_{\mathrm{sys},i}^{-1}\,\boldsymbol{g}_{\mathrm{sys},i}.
\label{eq30}
\end{equation}
Finally, the attitude can be updated on the manifold via a left-multiplicative exponential map,
\begin{equation}
\widehat{\boldsymbol{R}}_{i+1}=\exp\big([\delta\boldsymbol{\phi}_i]_{\times}\big)\,\widehat{\boldsymbol{R}}_{i},
\label{eq31}
\end{equation}
and the iteration stops when $\|\delta\boldsymbol{\phi}_i\|\le\varepsilon$, where $\varepsilon$ is a predefined convergence threshold. The complete algorithm for short-range DoA-based AD is summarized in \textbf{Algorithm}.

\begin{algorithm}[!t]
\renewcommand{\thealgorithm}{}
\caption{Manifold Gauss--Newton for Short-Range DoA-based AD}
\begin{algorithmic}[1]
\Require Observation vectors $\{\tilde{\boldsymbol{v}}_k^{n}\}_{k=1}^K$, reference vectors $\{\tilde{\boldsymbol{v}}_k^{b}\}_{k=1}^K$, with corresponding covariances $\{\boldsymbol{\Sigma}_{\boldsymbol{v}_k^n}\}_{k=1}^K$ and $\{\boldsymbol{\Sigma}_{\boldsymbol{v}_k^b}\}_{k=1}^K$, convergence threshold $\varepsilon$.
\Ensure Attitude estimate $\widehat{\boldsymbol{R}}$.
\State Initialize $\widehat{\boldsymbol{R}}_0$ (e.g., identity matrix or through the Wahba solution in \eqref{eq3}-\eqref{eq6}), set $i=0$.
\Repeat
    \For{$k=1$ to $K$}
        \State Compute $\hat{\boldsymbol{v}}_{k,i}^{n}$, $\boldsymbol{z}_{k,i}$, $\boldsymbol{J}_{k,i}$, $\breve{\boldsymbol{\Sigma}}_{\boldsymbol{v}_{k,i}}$, $\boldsymbol{W}_{k,i}$ using \eqref{eq27}.
    \EndFor
    \State Compute $\boldsymbol{F}_{\mathrm{sys},i}$, $\boldsymbol{g}_{\mathrm{sys},i}$ using \eqref{eq29}.
    \State Solve for $\delta\boldsymbol{\phi}_i$ using \eqref{eq30}.
    \State Update $\widehat{\boldsymbol{R}}_{i+1}$ using \eqref{eq31}.
    \State $i \leftarrow i+1$.
\Until{$\|\delta\boldsymbol{\phi}_i\|\le\varepsilon$}
\State \Return $\widehat{\boldsymbol{R}}=\widehat{\boldsymbol{R}}_i$.
\end{algorithmic}
\end{algorithm}

\subsection{Hessian-matching based weighting strategies} \label{sec:weighting_strategy}

The proposed manifold Gauss-Newton method effectively accounts for the total DV errors and addresses the anisotropic DV noise issue in short-range DoA-based AD. However, it requires multiple iterations and suitable regularization for the covariance matrix of DV errors, leading to increased computational complexity and reduced robustness compared to closed-form Wahba solutions. To bridge this gap, we next seek a closed-form weighting strategy that approximates the TLS solution while retaining the Wahba structure.

The Hessian matrix describes the local curvature of the objective and therefore quantifies how strongly each measurement constrains the unknown attitude in a neighborhood of the optimum. Although the mappings from DoA/position estimates to unit DVs involve nonlinear transformations and thus generally induce non-Gaussian errors, in the small-error regime the first-order linearization in \eqref{eq24} implies that the total DV error can be well approximated as Gaussian. Under this approximation, the TLS-consistent covariance-weighted objective in \eqref{eq26} is the negative log-likelihood of the attitude perturbation $\boldsymbol{\phi}$. Consequently, the associated Hessian coincides with the \emph{Fisher information matrix}, endowing the Hessian with a direct ``information content'' interpretation. From this perspective, Hessian matching can be interpreted as an information-matching principle: we choose scalar weights so that the Wahba objective reproduces, in the desired sense, the local information implied by the TLS model.

For a generic least-squares objective $J(\boldsymbol{\phi})$, its Hessian matrix can be defined as
\begin{equation}
\boldsymbol{H}(\boldsymbol{\phi})\ \triangleq\ \nabla^2_{\boldsymbol{\phi}} J(\boldsymbol{\phi})
\label{eq32}
\end{equation}
Therefore, the TLS-consistent objective in \eqref{eq26} induces a Hessian of the form
\begin{equation}
\boldsymbol{H}_{\mathrm{TLS}}\ \approx\ \sum_{k=1}^{K} \boldsymbol{J}_k^{\transpose}\,\breve{\boldsymbol{\Sigma}}_{\boldsymbol{v}_k}^{-1}\,\boldsymbol{J}_k,
\label{eq33}
\end{equation}
where $\boldsymbol{J}_k$ is the Jacobian defined in \eqref{eq24} and $\breve{\boldsymbol{\Sigma}}_{\boldsymbol{v}_k}$ is the regularized total DV error covariance in \eqref{eq17}. 

Similar to \eqref{eq26}, the scalar-weighted Wahba formulation in \eqref{eq2} can also be expressed in the difference-residual form as
\begin{equation}
J_{\mathrm{Wahba}}(\boldsymbol{\phi})
=\frac{1}{2}\sum_{k=1}^{K} w_k(\boldsymbol{z}_k - \boldsymbol{J}_k \boldsymbol{\phi})^{\transpose}(\boldsymbol{z}_k - \boldsymbol{J}_k \boldsymbol{\phi}).
\label{eq34}
\end{equation}
which induces a Hessian of the form:
\begin{equation}
\boldsymbol{H}_{\mathrm{Wahba}}\ \approx\ \sum_{k=1}^{K} w_k\,\boldsymbol{J}_{k}^{\transpose}\boldsymbol{J}_{k}.
\label{eq35}
\end{equation}

For \eqref{eq35}, using the identity $[\boldsymbol{v}]_{\times}^{2}=\boldsymbol{v}\boldsymbol{v}^{\transpose}-\|\boldsymbol{v}\|^{2}\boldsymbol{I}$, we obtain for the unit vector $\boldsymbol{v}_k^{n}$ that
\begin{equation}
\boldsymbol{J}_{k}^{\transpose}\boldsymbol{J}_{k}
=[\hat{\boldsymbol{v}}_k^{n}]_{\times}^{\transpose}[\hat{\boldsymbol{v}}_k^{n}]_{\times}
=-[\hat{\boldsymbol{v}}_k^{n}]_{\times}^{2}
=\boldsymbol{I}-\hat{\boldsymbol{v}}_k^{n}(\hat{\boldsymbol{v}}_k^{n})^{\transpose},
\label{eq36}
\end{equation}
which is a rank-2 projector onto the tangent plane orthogonal to $\hat{\boldsymbol{v}}_k^{n}$. This reveals that a single vector pair provides no information along its own axis, and therefore the rotation about that vector is unobservable.

Comparing \eqref{eq33} and \eqref{eq35} further clarifies the role of scalar weighting in the Wahba formulation. In the TLS-induced Hessian $\boldsymbol{H}_{\mathrm{TLS}}$, $\boldsymbol{J}_k$ captures the measurement geometry, while $\breve{\boldsymbol{\Sigma}}_{\boldsymbol{v}_k}^{-1}$ encodes the uncertainty of the $k$-th vector pair. In contrast, the Wahba's Hessian $\boldsymbol{H}_{\mathrm{Wahba}}$ contains only the geometric term $\boldsymbol{J}_k^{\transpose}\boldsymbol{J}_k$ and therefore cannot directly reflect heterogeneous uncertainty across measurements. This motivates Hessian matching: by properly choosing the scalar weights $\{w_k\}$, we aim to compensate for the missing uncertainty term and make $\boldsymbol{H}_{\mathrm{Wahba}}$ approximate $\boldsymbol{H}_{\mathrm{TLS}}$ in the sense of interest, so that the closed-form Wahba solver preserves, as much as possible, the local curvature/information content of the TLS model.

\paragraph*{Strategy A (full-attitude trace matching).}
A natural ``global'' matching criterion is to match the total curvature (trace) contributed by each vector pair,
\begin{equation}
\trace \big(w_k\,\boldsymbol{J}_{k}^{\transpose}\boldsymbol{J}_{k}\big)
\ \approx\ \trace \big(\boldsymbol{J}_k^{\transpose}\,\breve{\boldsymbol{\Sigma}}_{\boldsymbol{v}_k}^{-1}\,\boldsymbol{J}_k\big).
\label{eq37}
\end{equation}
Since $\trace(\boldsymbol{J}_{k}^{\transpose}\boldsymbol{J}_{k})=\trace(\boldsymbol{I}-\hat{\boldsymbol{v}}_k^{n}(\hat{\boldsymbol{v}}_k^{n})^{\transpose})=2$, and by applying the cyclic property of the trace operator, we have
\begin{equation}
\trace \big(\boldsymbol{J}_k^{\transpose}\,\breve{\boldsymbol{\Sigma}}_{\boldsymbol{v}_k}^{-1}\,\boldsymbol{J}_k\big)
=\trace\big(\breve{\boldsymbol{\Sigma}}_{\boldsymbol{v}_k}^{-1}\,(\boldsymbol{I}-\hat{\boldsymbol{v}}_k^{n}(\hat{\boldsymbol{v}}_k^{n})^{\transpose})\big) \approx \trace\big( \breve{\boldsymbol{\Sigma}}_{\boldsymbol{v}_k}^{-1} \big)
\label{eq38}
\end{equation}
The second approximate equality holds because $\hat{\boldsymbol{v}}_k^{n}(\hat{\boldsymbol{v}}_k^{n})^{\transpose}$ is a rank-1 matrix with eigenvalues 1 and 0, and $\boldsymbol{\Sigma}_{\boldsymbol{v}_k}^{-1}$ is orthogonal to it (since the noise lies in the tangent plane). Therefore, $\trace\big(\boldsymbol{\Sigma}_{\boldsymbol{v}_k}^{-1}\,\hat{\boldsymbol{v}}_k^{n}(\hat{\boldsymbol{v}}_k^{n})^{\transpose}\big)\approx0$. This yields the scalar weight
\begin{equation}
w_k^{(\boldsymbol{\phi})}=\frac{1}{2}\, \trace\big( \breve{\boldsymbol{\Sigma}}_{\boldsymbol{v}_k}^{-1} \big).
\label{eq39}
\end{equation}
where the superscript $(\boldsymbol{\phi})$ indicates that this weight is derived from full attitude trace matching, which accouts for the overall misalignment error. However, since the matrix $\boldsymbol{\Sigma}_{\boldsymbol{v}_k}$ is typically rank-deficient, even the regularized version $\breve{\boldsymbol{\Sigma}}_{\boldsymbol{v}_k}$ may still be ill-conditioned, leading to numerical instability in computing its inverse. We can take a step back and adopt a simpler and more numerically stable approach to compute this weight as\footnote{
Since $\boldsymbol{\Sigma}_{\boldsymbol{v}_k}$ is rank-deficient, we assume that it has two non-zero eigenvalues $\lambda_1$ and $\lambda_2$, whose corresponding eigenvectors span the tangent plane orthogonal to $\hat{\boldsymbol{v}}_k^{n}$. The regularized covariance $\breve{\boldsymbol{\Sigma}}_{\boldsymbol{v}_k}$ shares similar eigenvalues. In this way, $w_1 = \tfrac{1}{2} \trace(\breve{\boldsymbol{\Sigma}}_{\boldsymbol{v}_k}^{-1})\approx \tfrac{1}{2} \left( \tfrac{1}{\lambda_1} + \tfrac{1}{\lambda_2} \right)$, while $w_2 = \tfrac{2}{\trace\big( \boldsymbol{\Sigma}_{\boldsymbol{v}_k} \big)} = \tfrac{2}{\lambda_1 + \lambda_2}$. It is easy to find that $w_1\geq w_2$ by the arithmetic-harmonic mean inequality and they are equal if and only if $\lambda_1 = \lambda_2$. If the noise is not highly anisotropic, $w_2$ will be a good approximation of $w_1$; if the noise is highly anisotropic, then $w_1$ may overemphasize the less noisy direction and underemphasize the more noisy direction, while $w_2$ provides a more conservative average weighting, which may be more robust and preferred in practice. Moreover, dividing by a constant factor of 2 does not affect the solution to the Wahba problem, so we adopt the simpler expression in \eqref{eq40}.
}:
\begin{equation}
    w_k^{(\boldsymbol{\phi})} = \frac{1}{\trace\big( \breve{\boldsymbol{\Sigma}}_{\boldsymbol{v}_k} \big)} \approx \frac{1}{\trace\big( {\boldsymbol{\Sigma}}_{\boldsymbol{v}_k^b} \big) + \trace\big({\boldsymbol{\Sigma}}_{\boldsymbol{v}_k^n} \big)}
    \label{eq40}
\end{equation}
The proposed weight \eqref{eq40} explicitly incorporates both body-frame DV uncertainty (DoA-induced) and navigation-frame DV uncertainty (position-induced). In the long-range case (such as GNSS scenarios), $\trace(\breve{\boldsymbol{\Sigma}}_{\boldsymbol{v}_k^n})\approx 0$, and \eqref{eq40} naturally reduces to the DoA-only weighting form $w_k =  1/\trace({\boldsymbol{\Sigma}}_{\boldsymbol{v}_k^b})$ in \cite{wang2025attitude}.

\paragraph*{Strategy B (Direction-of-interest Hessian-matching).}
Strategy~A matches the \emph{overall} tangent-plane information via trace and therefore yields an ``average'' scalar weight in \eqref{eq39} and \eqref{eq40}. If one is instead interested in preserving the curvature/information along a specific rotation direction, we may match the Hessian only along that direction. Let $\boldsymbol{d}\in\mathbb{R}^3$ be a user-specified unit DOI expressed in the navigation frame. The directional curvature contributed by the $k$-th DV under the TLS model is defined as
\begin{equation}
K_{\mathrm{TLS},k}(\boldsymbol{d})\ \triangleq\ \boldsymbol{d}^{\transpose}\big(\boldsymbol{J}_k^{\transpose}\,\breve{\boldsymbol{\Sigma}}_{\boldsymbol{v}_k}^{-1}\,\boldsymbol{J}_k\big)\boldsymbol{d}.
\label{eq41}
\end{equation}
On the Wahba side, the corresponding directional curvature is
\begin{equation}
K_{\mathrm{Wahba},k}(\boldsymbol{d})\ \triangleq\ \boldsymbol{d}^{\transpose}\big(w_k\,\boldsymbol{J}_k^{\transpose}\boldsymbol{J}_k\big)\boldsymbol{d}
=w_k\,\|\hat{\boldsymbol{v}}_k^{n}\times\boldsymbol{d}\|^2.
\label{eq42}
\end{equation}
Define the tangent vector and its unit version as
\begin{equation}
\boldsymbol{u}_{\mathrm{tan},k}(\boldsymbol{d})\triangleq\hat{\boldsymbol{v}}_k^{n}\times\boldsymbol{d},\quad
\boldsymbol{u}_{\mathrm{unit},k}(\boldsymbol{d})\triangleq\frac{\boldsymbol{u}_{\mathrm{tan},k}(\boldsymbol{d})}{\|\boldsymbol{u}_{\mathrm{tan},k}(\boldsymbol{d})\|},
\label{eq43}
\end{equation}
when $\|\boldsymbol{u}_{\mathrm{tan},k}(\boldsymbol{d})\|\neq 0$. Matching $K_{\mathrm{Wahba},k}(\boldsymbol{d})=K_{\mathrm{TLS},k}(\boldsymbol{d})$ yields the DOI Hessian-matching weight
\begin{equation}
w_k^{(\boldsymbol{d})}
=\frac{K_{\mathrm{TLS},k}(\boldsymbol{d})}{\|\hat{\boldsymbol{v}}_k^{n}\times\boldsymbol{d}\|^2}
=\boldsymbol{u}_{\mathrm{unit},k}(\boldsymbol{d})^{\transpose}\,\breve{\boldsymbol{\Sigma}}_{\boldsymbol{v}_k}^{-1}\,\boldsymbol{u}_{\mathrm{unit},k}(\boldsymbol{d}).
\label{eq44}
\end{equation}
For the same reason as the simplification from \eqref{eq39} to \eqref{eq40} in Strategy~A, \eqref{eq44} can be further approximated as
\begin{equation}
w_k^{(\boldsymbol{d})} = \frac{1}{\boldsymbol{u}_{\mathrm{unit},k}(\boldsymbol{d})^{\transpose}\,{\boldsymbol{\Sigma}}_{\boldsymbol{v}_k}\,\boldsymbol{u}_{\mathrm{unit},k}(\boldsymbol{d})}.
\label{eq45}
\end{equation}

Specifically, if one is particularly concerned with the misalignment components $\phi_E$, $\phi_N$, and $\phi_U$, we can set $\boldsymbol{d}=\boldsymbol{e}_1=[1,0,0]^{\transpose}$, $\boldsymbol{d}=\boldsymbol{e}_2=[0,1,0]^{\transpose}$, and $\boldsymbol{d}=\boldsymbol{e}_3=[0,0,1]^{\transpose}$ to obtain the corresponding DOI Hessian-matching weights
\begin{equation}
    \begin{gathered}
        w_k^{(\phi_E)} = \frac{1}{\boldsymbol{u}_{\mathrm{unit},k}(\boldsymbol{e}_1)^{\transpose}\,{\boldsymbol{\Sigma}}_{\boldsymbol{v}_k}\,\boldsymbol{u}_{\mathrm{unit},k}(\boldsymbol{e}_1)}, \\
        w_k^{(\phi_N)} = \frac{1}{\boldsymbol{u}_{\mathrm{unit},k}(\boldsymbol{e}_2)^{\transpose}\,{\boldsymbol{\Sigma}}_{\boldsymbol{v}_k}\,\boldsymbol{u}_{\mathrm{unit},k}(\boldsymbol{e}_2)}, \\
        w_k^{(\phi_U)} = \frac{1}{\boldsymbol{u}_{\mathrm{unit},k}(\boldsymbol{e}_3)^{\transpose}\,{\boldsymbol{\Sigma}}_{\boldsymbol{v}_k}\,\boldsymbol{u}_{\mathrm{unit},k}(\boldsymbol{e}_3)}
    \end{gathered}
    \label{eq46}
\end{equation}
where the superscripts $(\phi_E)$, $(\phi_N)$, and $(\phi_U)$ indicate that the weights are derived by matching the directional curvature along the east, north, and up axes in the ENU navigation frame, respectively, aiming to reduce the corresponding misalignment errors. This further explains our choice of treating the navigation-frame DVs as observation vectors and the body-frame DVs as reference vectors: it enables the above DOI Hessian-matching strategies to be implemented entirely in the navigation frame, where the misalignment errors of interest admit a simple unit-vector representation.

\section{IMU-Assisted AD with Gravity DV Augmentation} \label{method2}
This section develops IMU-assisted AD methods tailored to static initialization. The key idea is to incorporate IMU-derived gravity information into the same DV-alignment solver as the DoA-derived DV pairs, rather than using gravity only for a standalone leveling step or as an external prior. Concretely, we treat the gravity direction as an additional DV pair and augment it into the DoA-based DV alignment.

As reviewed in Section~\ref{sec:leveling}, when the vehicle is static, a unit gravity DV observation in the body frame and its known counterpart in the navigation frame can be constructed via \eqref{eq7}. In the noise-free case, $\boldsymbol{v}_{\boldsymbol{g}}^{n}=\boldsymbol{R}\boldsymbol{v}_{\boldsymbol{g}}^{b}$. As discussed in Section~\ref{problem-formulation}, after temporal averaging the dominant residual accelerometer error is the turn-on bias, modeled as $\Delta\boldsymbol{f}^b = \boldsymbol{b}_a$ with $\boldsymbol{b}_a \sim \mathcal{N}(\boldsymbol{0},\sigma_{b_a}^2\boldsymbol{I})$. Linearizing the unit normalization in the definition of $\tilde{\boldsymbol{v}}_{\boldsymbol{g}}^{b}$ in \eqref{eq7} yields
\begin{equation}
    \Delta\boldsymbol{v}_{\boldsymbol{g}}^{b}
    \approx \frac{1}{g}\,\boldsymbol{P}_{\tilde{\boldsymbol{v}}_{\boldsymbol{g}}^{b}}^{\perp}\,\Delta\boldsymbol{f}^b,
    \quad
    \boldsymbol{P}_{\tilde{\boldsymbol{v}}_{\boldsymbol{g}}^{b}}^{\perp}\triangleq\boldsymbol{I}-\tilde{\boldsymbol{v}}_{\boldsymbol{g}}^{b}(\tilde{\boldsymbol{v}}_{\boldsymbol{g}}^{b})^{\transpose}.
    \label{eq47}
\end{equation}
Consequently, the gravity-DV error covariance is derived by
\begin{equation}
    \boldsymbol{\Sigma}_{\boldsymbol{v}_{\boldsymbol{g}}^{b}}
    \triangleq \mathbb{E} \left[\Delta\boldsymbol{v}_{\boldsymbol{g}}^{b}(\Delta\boldsymbol{v}_{\boldsymbol{g}}^{b})^{\transpose}\right]
    \approx \frac{\sigma_{b_a}^2}{g^2}\,\boldsymbol{P}_{\tilde{\boldsymbol{v}}_{\boldsymbol{g}}^{b}}^{\perp}.
    \label{eq48}
\end{equation}
Since $\boldsymbol{v}_{\boldsymbol{g}}^{n}$ is known, the observation-side uncertainty of the gravity DV is negligible and the dominant uncertainty lies in the body-frame estimate $\tilde{\boldsymbol{v}}_{\boldsymbol{g}}^{b}$. Then the IMU-derived gravity DV pair can be treated similarly to the DoA-derived DV pairs in Section~\ref{method1}, and be utilized in the DV-alignment framework in two ways:
\begin{itemize}
    \item As an additional DV pair in the TLS-consistent covariance-weighted formulation in Section \ref{sec:tls_formulation}
    \item As an additional DV pair in the closed-form scalar-weighted Wahba formulation in Section \ref{sec:weighting_strategy}
\end{itemize}
To distinguish the IMU-assisted formulations from the DoA-only counterparts, we call the former one as the IMU-assisted \emph{extended TLS} (ETLS) formulation and the latter one as the IMU-assisted \emph{extended Wahba} (EWahba) formulation. Since the ETLS formulation has no difference in structure from the TLS formulation in Section~\ref{sec:tls_formulation}, we focus on the IMU-assisted EWahba formulation in the following.

Typically, for the gravity DV pair, we can assign a weight based on the Hessian-matching principle in \eqref{eq40} as
\begin{equation}
    w_{\boldsymbol{g}} \triangleq \frac{1}{\trace (\boldsymbol{\Sigma}_{\boldsymbol{v}_{\boldsymbol{g}}^{b}})} \approx \frac{g^2}{2\sigma_{b_a}^2},
    \label{eq49}
\end{equation}
where the last approximation uses $\trace(\boldsymbol{P}_{\tilde{\boldsymbol{v}}_{\boldsymbol{g}}^{b}}^{\perp})=2$. With the short-range DoA-based DV pairs $\{(\tilde{\boldsymbol{v}}_k^n,\tilde{\boldsymbol{v}}_k^b)\}_{k=1}^K$ and weights $\{w_k\}_{k=1}^K$ given by the proposed Hessian-matching based rule in \eqref{eq40}, we augment the DV pair set with the gravity DV pair in \eqref{eq7}. The IMU-assisted \emph{EWahba} objective is written as
\begin{equation}
\begin{split}
    J_{\mathrm{EWahba}}(\boldsymbol{R})
    = &\sum_{k=1}^{K} w_k \norm{\tilde{\boldsymbol{v}}_k^{n} - \boldsymbol{R} \tilde{\boldsymbol{v}}_k^{b}}^2
    + w_{\boldsymbol{g}} \norm{\boldsymbol{v}_{\boldsymbol{g}}^{n} - \boldsymbol{R} \tilde{\boldsymbol{v}}_{\boldsymbol{g}}^{b}}^2 \\
    = &\sum_{k=1}^{K+1} w_k \norm{\tilde{\boldsymbol{v}}_k^{n} - \boldsymbol{R} \tilde{\boldsymbol{v}}_k^{b}}^2,
\end{split}
    \label{eq50}
\end{equation}
where the augmented ``$(K{+}1)$-th'' pair is defined by
\begin{equation}
    w_{K+1} \triangleq w_{\boldsymbol{g}}, \quad 
    \tilde{\boldsymbol{v}}_{K+1}^{b} \triangleq \tilde{\boldsymbol{v}}_{\boldsymbol{g}}^{b}, \quad 
    \tilde{\boldsymbol{v}}_{K+1}^{n} \triangleq \boldsymbol{v}_{\boldsymbol{g}}^{n}.
\label{eq51}
\end{equation}
Minimizing \eqref{eq50} over $\boldsymbol{R}\in\mathrm{SO}(3)$ is again a weighted Wahba problem and can therefore be solved in closed form using the SVD procedure in Section~\ref{sec:wahba}.

Furthermore, the gravity DV provided by an IMU is typically much more accurate than the direction vectors obtained from short-range DoA measurements, and thus offers strong constraints on horizontal attitude (roll and pitch) but relatively weak constraints on yaw. Therefore, the short-range DoA-derived DV pairs can be primarily exploited to constrain yaw, so as to combine the complementary strengths of IMU and DoA. In this case, the weights of the DoA-derived DV pairs in \eqref{eq50} can adopt the DOI Hessian-matching form in \eqref{eq46} by setting $\boldsymbol{d}=\boldsymbol{e}_3=[0,0,1]^{\transpose}$. We shall see in Section~\ref{sim2} that this strategy effectively improves yaw estimation while maintaining good roll/pitch performance.

\section{Simulation Results} \label{simulation}
This section evaluates the proposed short-range DoA-based AD and IMU-assisted AD methods via Monte Carlo simulations. We first introduce the simulation setup and performance metrics. We then evaluate (i) the advantage of accounting for anisotropic two-sided DV uncertainties through covariance-weighted TLS versus scalar-weighted Wahba, (ii) the effectiveness of the proposed Hessian-matching weighting strategies in the closed-form Wahba solver, and (iii) the gain and robustness improvement brought by gravity-DV augmentation under different IMU qualities and anchor availability.

\subsection{Simulation Setup}
Unless otherwise specified, throughout the simulations, the vehicle is assumed to be located at the origin, $\boldsymbol{p}_0^{n}=[0,0,0]^{\transpose}$, with its position estimate corrupted by zero-mean Gaussian noise characterized by a standard deviation of $\sigma_{p_0}=0.25$ m. The true attitude is set to zero roll, zero pitch, and zero yaw (i.e., $\boldsymbol{R}=\boldsymbol{I}$), ensuring that the navigation frame is perfectly aligned with the body frame. Under this setup, the reported misalignment errors directly correspond to the errors in roll, pitch, and yaw. Each experiment is evaluated over $N_{\mathrm{sim}}=10{,}000$ Monte Carlo trials

\subsubsection{Error Metrics}
Since Euler angles do not form a vector space in the strict mathematical sense, attitude errors cannot be rigorously quantified using simple arithmetic differences. To address this, we adopt the widely used Phi-angle misalignment vector, $\boldsymbol{\phi}=[\phi_E,\phi_N,\phi_U]^{\transpose}$, expressed in the ENU navigation frame, as a measure to characterize attitude errors \cite{niu2021wheel}. The Phi-angle misalignment vector provides an intuitive representation of the rotational misalignment between the estimated and true navigation frames. Specifically, we report the component-wise misalignment errors $\{|\phi_E|, |\phi_N|, |\phi_U|\}$, corresponding to the east, north, and up directions, respectively, along with the overall attitude error magnitude $\|\boldsymbol{\phi}\|$.

The primary metric for evaluating attitude errors is the Monte Carlo root-mean-square error (RMSE), defined as:
\begin{equation}
\begin{gathered}
\mathrm{RMSE}_{\boldsymbol{\phi}} \triangleq \sqrt{\mathbb{E}[\|\boldsymbol{\phi}\|^2]}, \\
\mathrm{RMSE}_{\boldsymbol{\phi}_D} \triangleq \sqrt{\mathbb{E}[\|\boldsymbol{\phi}_D\|^2]}, \quad D = E, N, U,
\end{gathered}
\label{eq52}
\end{equation}
where $\mathrm{RMSE}_{\boldsymbol{\phi}}$ quantifies the overall attitude error, and $\mathrm{RMSE}_{\boldsymbol{\phi}_D}$ represents the error along a specific direction $D$ (east, north, or up). 
In addition, for specific experiments, we also report the covariance-predicted root-mean-trace error (RMTE), which is defined as:
\begin{equation}
\mathrm{RMTE}\triangleq \sqrt{\trace(\boldsymbol{\Sigma}_{\boldsymbol{\phi}})},
\label{eq53}
\end{equation}
where $\boldsymbol{\Sigma}_{\boldsymbol{\phi}}$ denotes the attitude error covariance derived from the weighted-Wahba covariance analysis \cite{markley1988attitude}. 

\subsubsection{Array Configuration}
Although the proposed formulations are not restricted to any specific radio technology or array configuration, we adopt a three-element planar UWB array in our simulations for two key reasons. First, UWB technology is widely utilized in unmanned systems and indoor navigation due to its high temporal resolution, which facilitates reliable short-range ranging and bearing estimation \cite{IEEE802154z,dotlic2017angle}. Second, the compact three-element planar array inherently exhibits direction-dependent DoA accuracy with anisotropic noise characteristics, making it a representative and nontrivial testbed for evaluating the proposed weighting strategy. The DoA error statistics for the selected UWB array are detailed in \hyperref[appendix]{Appendix}, and the main array parameters used for generating AoA measurements are summarized in Table~\ref{tab1}.
\begin{table}[!t]
    \centering
    \caption{UWB Array Configuration Parameters}
    \label{tab1}
    \begin{tabular}{
        >{\centering\arraybackslash}p{1cm}
        >{\centering\arraybackslash}p{1.5cm}
        >{\centering\arraybackslash}p{1.2cm}
        >{\centering\arraybackslash}p{1cm}
        >{\centering\arraybackslash}p{1cm}
    }
    \toprule
    Channel &$f_c$  &$\lambda$  &$d$  &$\sigma_{\Phi}$   \\
    \midrule
     5     & 6489.6 MHz   &4.62 cm   &0.95$\frac{\lambda}{2}$ &$5^{\circ}$ \cite{QorvoDW3000}  \\
    \bottomrule
    \end{tabular}
\end{table}

\subsubsection{Anchor Configurations}
Two anchor configurations are considered in different experiments.

\begin{figure*}
\centering
\subfloat[top view]{\includegraphics[width=0.4\linewidth]{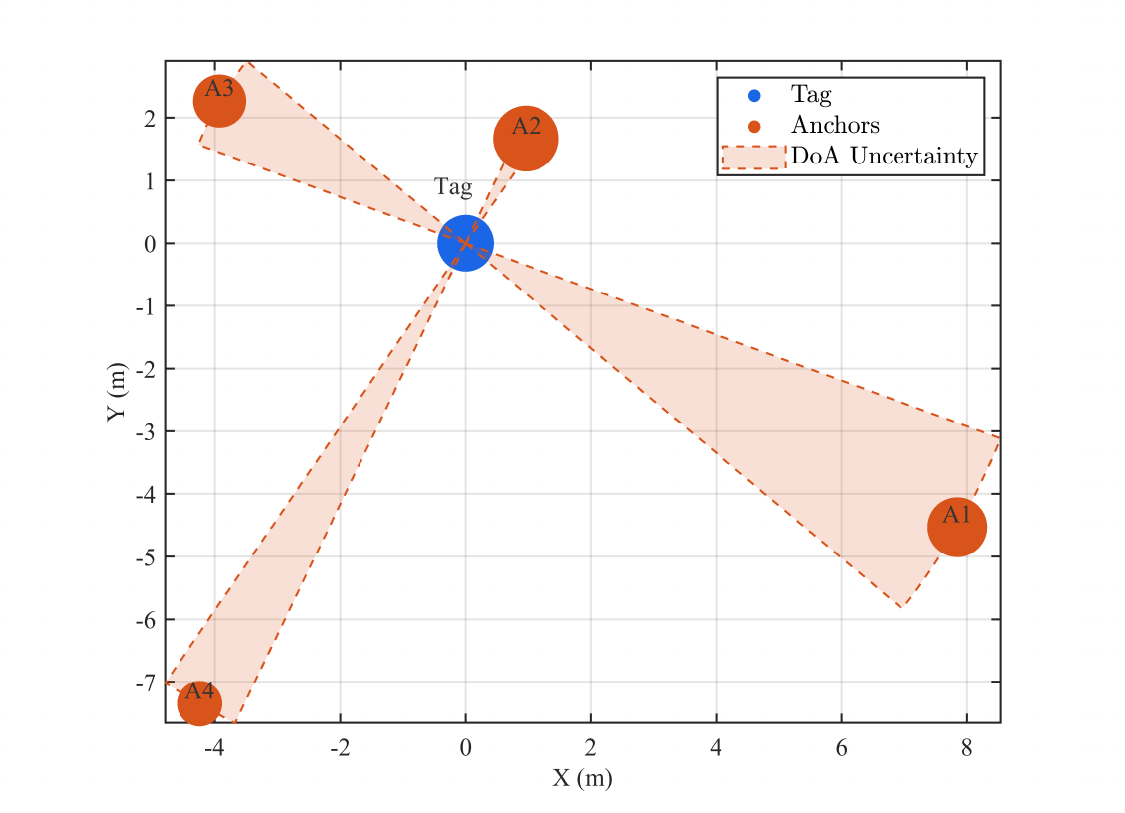}
\label{fig3_1}}
\hfil
\subfloat[side view]{\includegraphics[width=0.4\linewidth]{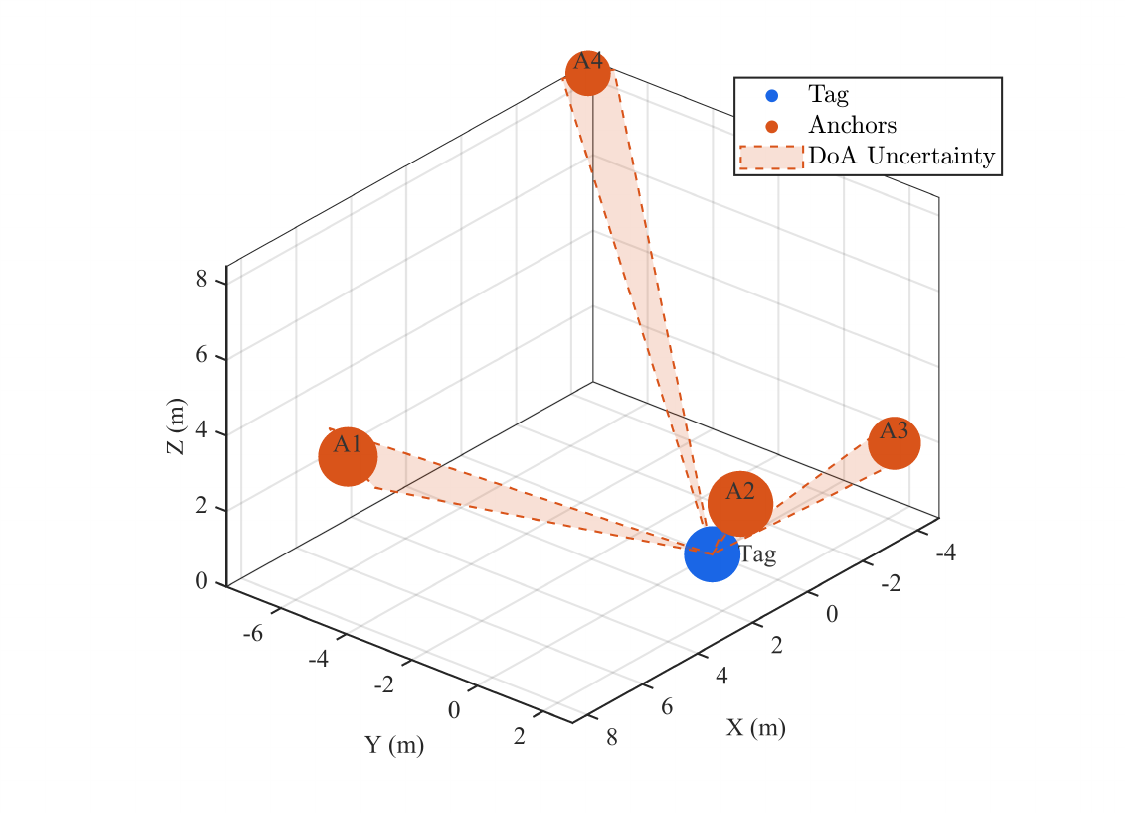}
\label{fig3_2}}
\caption{Deterministic anchor configuration. The vehicle equipped with a three-element array is located at the origin, while four anchors are distributed in 3D space with different azimuths, elevations, distances, and position uncertainties.}
\label{fig3}
\end{figure*}

\emph{Deterministic configuration}: four anchors are placed in 3D space as illustrated in Fig.~\ref{fig3}, with their azimuth, elevation, distance, and position uncertainty summarized in Table~\ref{tab2}. This configuration is used to highlight how geometry and position uncertainty can make navigation-frame DV errors non-negligible in short-range settings.

\emph{Random configuration}: to mitigate potential bias from a specific layout and to evaluate average performance across diverse geometries, anchor azimuth angles are drawn uniformly from $(-180^{\circ},180^{\circ}]$, elevations from $[15^{\circ},75^{\circ}]$, distances from $[\tfrac{1}{2}d_{\mathrm{avg}},\tfrac{3}{2}d_{\mathrm{avg}}]$ (with $d_{\mathrm{avg}}=10$ m by default), and anchor position uncertainties $\sigma_{p_k}$ from $[0.1,0.5]$ m.

\begin{table}[!t]
    \centering
    \caption{Deterministic Anchor Configuration}
    \label{tab2}
    \begin{tabular}{
        >{\centering\arraybackslash}p{1.5cm}
        >{\centering\arraybackslash}p{0.8cm}
        >{\centering\arraybackslash}p{0.8cm}
        >{\centering\arraybackslash}p{1.2cm}
        >{\centering\arraybackslash}p{1.2cm}
    }
    \toprule
    Anchor ID  &$\alpha_k$ ($^{\circ}$)  &$\beta_k$ ($^{\circ}$)  &$d_{k;0}$ (m)  &$\sigma_{p_k}$ (m)  \\
    \midrule
     A1  &-30    &25   &10   &0.3  \\
     A2  &60     &50   &3    &0.4  \\
     A3  &150    &25   &5    &0.2  \\
     A4  &-120   &45   &12   &0.1  \\
    \bottomrule
    \end{tabular}
\end{table}

\subsubsection{IMU Model}
The IMU frame is assumed to be aligned with the array/body frame. We consider two representative IMU qualities: a near-tactical-grade IMU (ADIS16465) \cite{analog2020adis16465} and a low-cost MEMS IMU (MPU6050) \cite{MPU6000Datasheet}. The standard deviation of the accelerometer turn-on bias is set to $\sigma_{b_a}=1400\,\mu\text{g}$ and $\sigma_{b_a}=20\,\text{mg}$, respectively, which directly determines the accuracy of gravity-DV.

\subsection{Evaluation of Short-Range DoA-based AD} \label{sim1}

\subsubsection{Covariance-weighted TLS versus scalar-weighted Wahba}
We first evaluate the impact of modeling anisotropic DV error statistics by comparing two approaches: (i) the covariance-weighted TLS objective solved using the manifold Gauss--Newton method, and (ii) the closed-form scalar-weighted Wahba estimator employing the proposed Hessian-matching scalar weights (Strategy~A, \eqref{eq40}). Three noise scenarios are considered: an isotropic DV noise case, an anisotropic case based on the deterministic anchor configuration, and an anisotropic case dominated by DoA-induced anisotropy. To ensure a fair comparison, the parameters are adjusted such that the average trace of the DV covariance matrices remains approximately the same across all three cases, thereby maintaining comparable overall DV noise levels. It is important to note that the manifold Gauss--Newton method for solving the TLS problem requires regularization due to the potential singularity of the total DV error covariance matrix. The choice of the regularization parameter significantly affects the performance of attitude estimation. In this study, we employ a grid search to determine the optimal regularization parameter, ensuring the best achievable performance for the TLS-based approach.

\begin{figure}[!b]
    \centering
    \includegraphics[width=\linewidth]{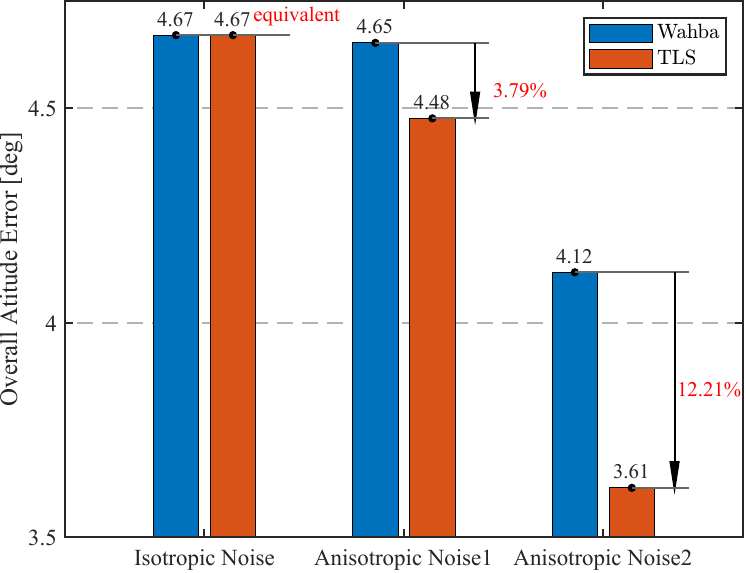}
    \caption{Overall attitude error comparison between the covariance-weighted TLS solver and the scalar-weighted Wahba solver under isotropic and anisotropic DV noise settings.}
    \label{fig4}
\end{figure}

Fig.~\ref{fig4} presents the overall attitude errors for different noise scenarios. As expected, under isotropic DV noise, the TLS and Wahba formulations yield identical performance, consistent with the theoretical result in \cite{chang2015total} that the weighted-Wahba formulation is a special case of TLS when noise is isotropic. However, under anisotropic DV noise, the TLS formulation, which explicitly accounts for full covariance information, achieves better accuracy than the scalar-weighted Wahba formulation. Specifically, the overall attitude errors are reduced by 3.79\% and 12.21\% in the Anisotropic1 and Anisotropic2 cases, respectively. The performance gap is attributed to the inability of the scalar-weighted Wahba formulation to accurately model the statistical characteristics of anisotropic DV noise. In the Anisotropic1 case, where position-induced navigation-frame DV errors dominate, the tangent-plane noise remains isotropic, allowing the Wahba formulation to perform reasonably well, resulting in a smaller performance gap. In contrast, in the Anisotropic2 case, where DoA-induced anisotropy dominates due to the inherent characteristics of the antenna array (e.g., significant differences in azimuth and elevation estimation accuracy, as shown in Fig.~\ref{fig11}), the performance gap becomes much more pronounced. 

A comprehensive comparison of the two formulations is summarized in Table \ref{tab3}. Generally, while the TLS formulation offers the highest accuracy across all scenarios, it comes with increased computational complexity (iterative nature of its solver) and reduced robustness (sensitive to regularization). In contrast, the scalar-weighted Wahba formulation provides a favorable trade-off between accuracy, computational efficiency, and robustness, making it a practical choice for real-time applications. Except for some extreme anisotropic cases, the Wahba formulation with suitable weights can achieve high accuracy close to that of TLS, so in the following sections, we will focus on evaluating different weighting strategies within the closed-form Wahba framework.

\begin{table}[!b]
\centering
\caption{Comparison of Two Attitude Determination Formulations \emph{TLS versus Wahba}}
\label{tab3}
\begin{tabular}{
    >{\centering\arraybackslash}m{3cm} 
    >{\centering\arraybackslash}m{2.5cm} 
    >{\centering\arraybackslash}m{1.5cm}
}
\toprule
\textbf{Item} & \textbf{TLS} & \textbf{Wahba} \\
\midrule
Accuracy (Isotropic)   & Highest & Highest \\
Accuracy (General)     & Highest & High \\
Accuracy (Anisotropic) & Highest & Modest \\
Complexity             & High    & Modest \\
Robustness             & Low     & High \\
Method                 & Iterative manifold Gauss--Newton & Closed-form SVD \\
\bottomrule
\end{tabular}
\end{table}

\subsubsection{Impact of weighting strategies in the closed-form Wahba solver}
We next evaluate the effect of different weighting strategies under the deterministic anchor configuration. We compare the following weighting schemes:
\begin{itemize}
    \item \textbf{Unweighted}: $\{w_k\}$ are uniform;
    \item \textbf{DoA weighted}: weights depend only on the body-frame (DoA-induced) DV uncertainty and ignore navigation-frame (position-induced) DV uncertainty \cite{wang2025attitude};
    \item \textbf{Hessian-matching weighted}: the proposed weighting scheme based on Hessian-matching (Strategy~A) in \eqref{eq40};
    \item \textbf{Optimal weighted} (benchmark): numerical minimization of the predicted $\trace(\boldsymbol{\Sigma}_{\boldsymbol{\phi}})$ over $\{w_k\}$ using the covariance derived in \cite{markley1988attitude}.
\end{itemize}

Table~\ref{tab4} reports the average DV error components and the resulting average weights assigned to each anchor. In short-range settings, the navigation-frame DV errors can be comparable to or even larger than the body-frame DV errors. Consequently, DoA-only weighting may overestimate the utility of an anchor whose DoA accuracy is high but whose position uncertainty produces a large navigation-frame DV error. For example, as highlighted in Table~\ref{tab4}, although Anchor 2 (A2) exhibits the smallest body-frame DV error (0.0027), which indicates excellent DoA accuracy, its resulting total DV error remains significant due to large position-induced errors (0.0534). The DoA-weighted strategy therefore assigns it a dominant weight (0.4078), whereas the proposed Hessian-matching weighting strategy, which explicitly accounts for both contributions, suppresses its weight to 0.0508. Comparison with the numerically obtained optimal weights shows that the proposed Hessian-matching based scheme closely approximates the optimized benchmark while remaining significantly simpler to compute, which provide an effective and practical alternative to the theoretically optimal weights.
\begin{table}[!b]
    \centering
    \caption{DV Error Components and Average Weights under Different Strategies (Deterministic Configuration)}
    \label{tab4}
    \begin{tabular}{
        >{\centering\arraybackslash}m{3cm}
        >{\centering\arraybackslash}m{0.8cm}
        >{\centering\arraybackslash}m{0.8cm}
        >{\centering\arraybackslash}m{0.8cm}
        >{\centering\arraybackslash}m{0.8cm}
    }
    \toprule
    Metric & A1 & A2 & A3 & A4 \\
    \midrule
    $\trace(\boldsymbol{\Sigma}_{\boldsymbol{v}_k^{b}})$      & 0.0248    & \textbf{0.0027}    & 0.0239 & 0.0029  \\
    $\trace(\boldsymbol{\Sigma}_{\boldsymbol{v}_k^{n}})$      & 0.0031   &  \textbf{0.0507}   & 0.0083  & 0.0010  \\
    $\trace(\boldsymbol{\Sigma}_{\boldsymbol{v}_k^{b}})+\trace(\boldsymbol{\Sigma}_{\boldsymbol{v}_k^{n}})$      & 0.0279   &  0.0534   & 0.0321  & \textbf{0.0039}  \\
    \midrule
    Unweighted $w_k$         & 0.2500   & 0.2500   & 0.2500  & 0.2500  \\
    DoA weighted $w_k$       & 0.1042   & \textbf{0.4078}   & 0.1047  & 0.3832  \\
    Hessian-matching weighted $w_k$      &  0.1769   & \textbf{0.0508}   & 0.1276  & \textbf{0.6447}  \\
    Optimal weighted $w_k$ & 0.1985   & \textbf{0.0406}   & 0.1457  & \textbf{0.6152}  \\
    \bottomrule
    \end{tabular}
\end{table}

\begin{figure}
    \centering
    \includegraphics[width=\linewidth]{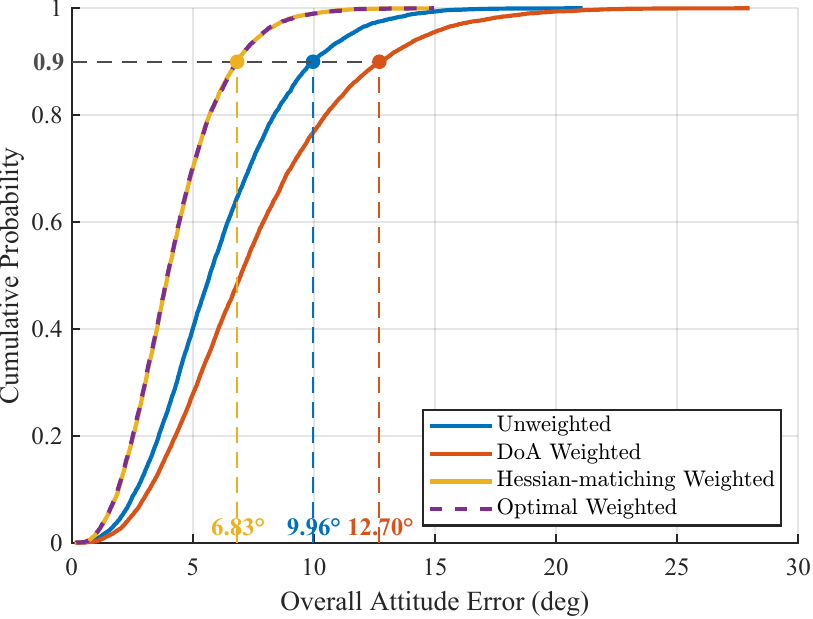}
    \caption{The CDF curve of the overall attitude error using different weighting strategies under the deterministic anchor configuration.}
    \label{fig5}
\end{figure}

The cumulative distribution function (CDF) of the overall attitude error is shown in Fig.~\ref{fig5}. The proposed Hessian-matching weighted scheme achieves performance nearly identical to the numerically optimized benchmark and outperforms both the DoA weighted and unweighted schemes. In particular, at the 90th percentile, the proposed scheme reduces the overall attitude error by approximately 46.21\% and 31.41\% compared to the DoA weighted and unweighted baselines. Moreover, the proposed weighting also shortens the tail of the error distribution, indicating improved robustness under unfavorable and extreme conditions.

\subsubsection{Performance versus anchor--vehicle distance}
\begin{figure}
    \centering
    \includegraphics[width=\linewidth]{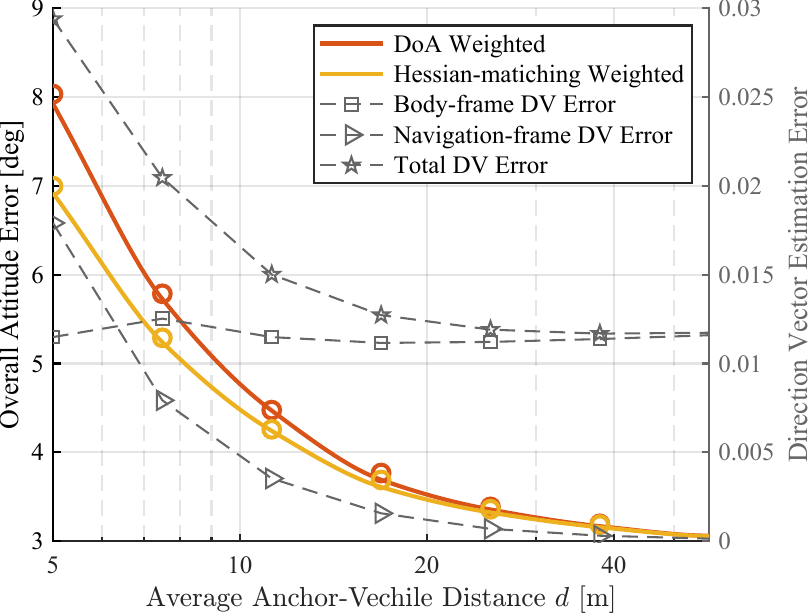}
    \caption{Attitude estimation performance versus the average anchor--vehicle distance under random anchor configurations (left: attitude error; right: DV error components).}
    \label{fig6}
\end{figure}

To demonstrate the distance dependence implied by the navigation-frame DV covariance in \eqref{eq14}, we evaluate performance versus the average anchor--vehicle distance under random anchor configurations. Fig.~\ref{fig6} reports (left) the resulting overall attitude errors and (right) the corresponding DV error components. As the distance increases, position-induced navigation-frame DV errors decrease rapidly, while DoA-induced body-frame DV errors remain nearly constant. Consistent with this trend, the gain of the proposed Hessian-matching weighting is most prominent in the short-range regime, where neglecting navigation-frame uncertainty leads to noticeable performance loss. As the distance increases toward the long-range regime, the proposed weight naturally approaches the DoA-only weighting form, and the two methods converge. Hence, the proposed scheme can be viewed as a generalization of \cite{wang2025attitude} to more general (including short-range) scenarios.

\subsubsection{DOI Hessian-matching weighting}

\begin{figure}
    \centering
    \includegraphics[width=\linewidth]{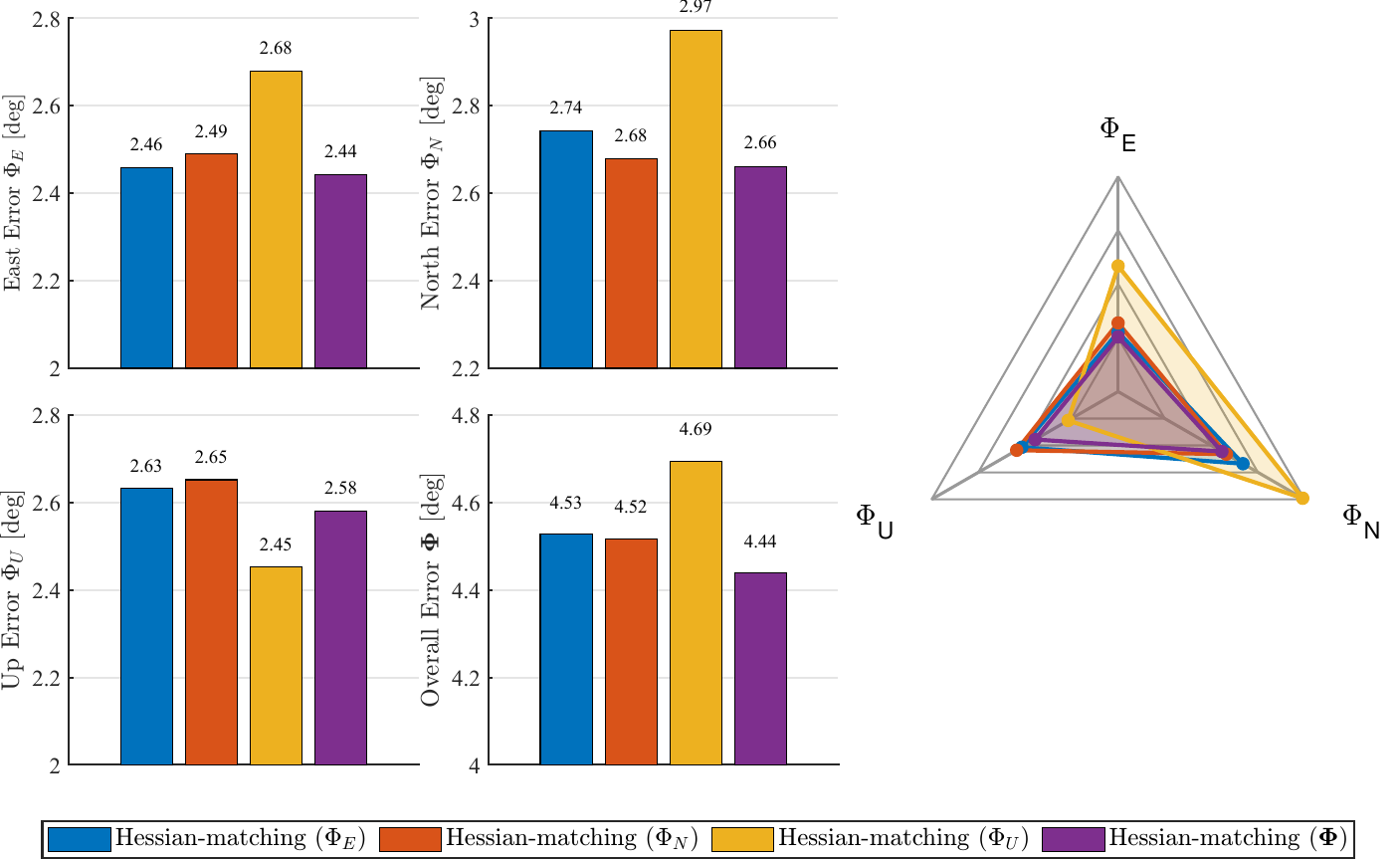}
    \caption{Component-wise attitude errors and overall attitude error under random anchor configurations using different DoI choices in Strategy~B of Hessian-matching based weighting scheme, compared with the full-attitude Strategy~A.}
    \label{fig7}
\end{figure}

Finally, we evaluate Strategy~B in Hessian-matching based weighting under random anchor configurations. By selecting $\boldsymbol{d}=\boldsymbol{e}_1$, $\boldsymbol{e}_2$, and $\boldsymbol{e}_3$ in the ENU navigation frame, the resulting scalar weight emphasizes the local curvature/information along the corresponding misalignment component, namely $\phi_E$, $\phi_N$, and $\phi_U$, respectively. Fig.~\ref{fig7} compares the resulting component-wise misalignment errors as well as the overall attitude estimation error. The results indicate that, by selecting different preferred directions, the attitude estimation performance along the corresponding component can be further improved. Although the DOI Hessian-matching weights $\phi_E$ and $\phi_N$ do not outperform the overall-error–minimizing weighting strategy in terms of east- and north-axis misalignment errors—primarily due to the combined effects of anchor geometry, they nevertheless exhibit noticeably better performance than other directional weighting schemes not specified to its direction. In contrast, the Hessian-matching weight $\phi_U$ does not yield the lowest overall attitude error, but achieves a significant reduction in the up-axis misalignment error. These results demonstrate that Strategy~B provides a flexible mechanism to trade overall estimation accuracy for improved performance along a user-prioritized direction. This property becomes particularly useful when DoA measurements are fused with IMU-assisted information, as discussed in Section~\ref{sim2}.

\subsection{Evaluation of IMU-Assisted AD} \label{sim2}
We next evaluate the gain and robustness improvement brought by IMU-derived gravity-DV under different IMU qualities and anchor availability.

\subsubsection{Impact of IMU quality}
\begin{figure}[!b]
    \centering
    \includegraphics[width=\linewidth]{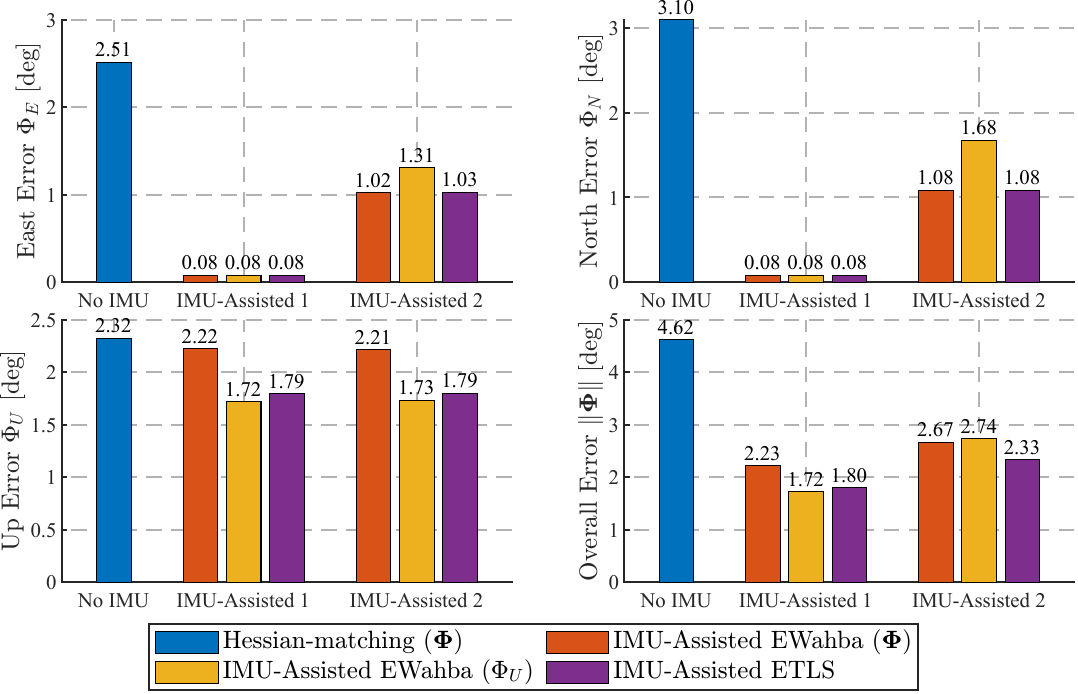}
    \caption{Component-wise attitude errors and overall attitude error under deterministic anchor configurations. ``IMU-Assisted 1'' and ``IMU-Assisted 2'' correspond to a near-tactical-grade and a low-cost IMU, respectively.}
    \label{fig8}
\end{figure}

First, we compare (i) the pure DoA-based estimator without IMU assistance, (ii) the IMU-assisted EWahba formulation, and (iii) the IMU-assisted ETLS formulation, which serves as a statistical performance reference. For the DoA-derived DV pairs, the full-attitude Hessian-matching based weighting strategy in \eqref{eq40} (Strategy~A) and the yaw-focused DOI weighting strategy (Strategy~B with $\boldsymbol{d}=\boldsymbol{e}_3$) are adopted to compare the DoA contribution along the yaw-related axis. The resulting misalignment errors are reported in Fig.~\ref{fig8}. As expected, for all IMU-assisted methods, augmenting the DV set with gravity-based information significantly reduces the horizontal-axis misalignment components $\Phi_E$ and $\Phi_N$, which are primarily associated with roll and pitch errors. This improvement arises because the gravity vector provides a strong and highly reliable constraint along these directions. In contrast, the improvement in the up-axis misalignment component $\Phi_U$ is limited compared with the other two components. 

In scenarios with relatively high IMU accuracy, Strategy~A does not explicitly prioritize yaw-related information but instead allocates weights according to the intrinsic noise characteristics of the DoA measurements. As a result, a substantial portion of the DoA information is implicitly used to further optimize the east and north components. However, when accurate gravity information is already available, this additional allocation provides limited benefit for the east and north axes and, in turn, restricts the potential improvement along the up axis. In contrast, Strategy~B explicitly concentrates the DoA-derived information on the up direction, which is not observable from gravity measurements. By exploiting this complementary property, Strategy~B achieves a more effective utilization of the DoA information and leads to a significant reduction in the up-axis misalignment error, with $\Phi_U$ further decreasing by approximately 22.5\% compared to Strategy~A.

When the IMU accuracy is lower, however, DoA measurements continue to provide meaningful information for reducing the east- and north-axis errors. Under such conditions, Strategy~A becomes more effective in improving the overall attitude estimation performance. These results indicate that, when incorporating gravity-derived DVs into a DV-alignment framework, the weighting strategy should be selected in accordance with both the IMU accuracy level and the specific attitude components of interest. Such an adaptive selection enables a more balanced and application-oriented exploitation of the complementary information provided by DoA and IMU measurements.

\subsubsection{Impact of anchor availability}
\begin{figure}
    \centering
    \includegraphics[width=\linewidth]{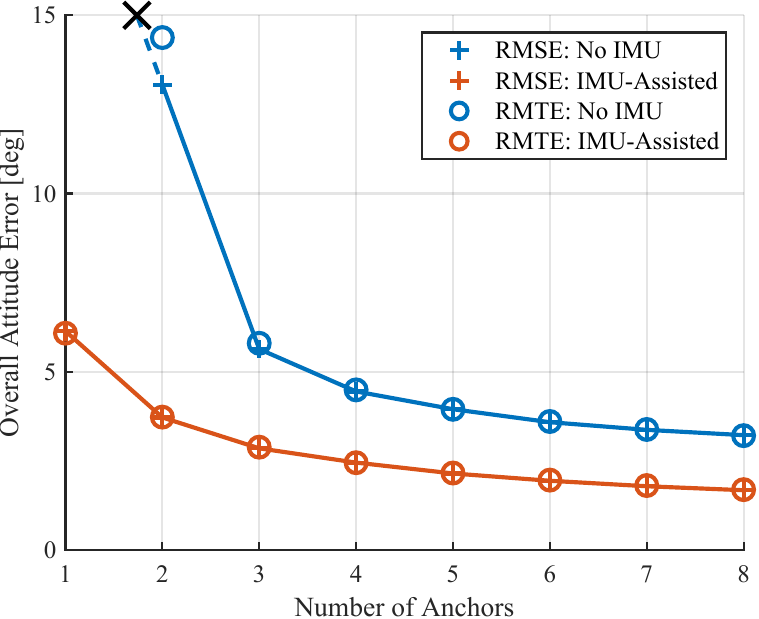}
    \caption{Overall attitude error as a function of the number of available anchors under random anchor configurations, comparing the pure DoA-based estimator and the IMU-assisted estimator.}
    \label{fig9}
\end{figure}

The number of available anchors also has a direct impact on AD performance, and we further investigate the robustness of the proposed methods with respect to anchor availability. Fig.~\ref{fig9} plots the overall attitude error versus anchor count, reporting both RMSE (Monte Carlo) and RMTE (covariance prediction). As the number of anchors increases, the performance of both the pure DoA-based estimator and the IMU-assisted estimator improves steadily, owing to the enhanced geometric diversity and measurement redundancy. For the same anchor count, the IMU-assisted method consistently achieves lower estimation errors, highlighting the benefit of incorporating the gravity-derived direction vector from the IMU. In particular, by providing an additional non-collinear reference direction, the IMU-assisted approach significantly improves solution availability and robustness in limited anchor availability scenarios. Moreover, the close agreement between the RMSE and RMTE indicates that the proposed covariance analysis accurately captures the estimation uncertainty, demonstrating the strong predictability and reliability of the proposed methods.

\section{Conclusion} \label{conclusion}
This paper studied AD for unmanned vehicles using short-range, DoA-capable wireless arrays, with optional IMU assistance for static initialization. We revisited short-range DoA-based AD from a TLS viewpoint by modeling the total DV error arising jointly from DoA estimation and position-induced uncertainties. Unlike the isotropic assumptions often adopted in the literature, the resulting DV errors are generally anisotropic and geometry dependent, motivating a covariance-weighted TLS formulation. We solved the corresponding covariance-weighted orthogonal Procrustes problem via a manifold Gauss--Newton method on $\mathrm{SO}(3)$ under general covariance structures. The results highlighted the importance of accurately modeling anisotropic DV uncertainties, as the proposed TLS formulation outperformed the classical scalar-weighted Wahba approach when such anisotropies were prominent.

To retain the efficiency and numerical robustness of the closed-form SVD-based Wahba solver, we further developed Hessian-matching based scalar weighting strategies that approximate the local Hessian geometry implied by the TLS model. The full-attitude strategy targets overall estimation accuracy, while the direction-of-interest strategy enables deliberate prioritization of a selected attitude component when application needs dictate. Simulation results showed that the Hessian-matching weighted estimator improves robustness compared with existing weighting baselines, especially in short-range settings where position-induced reference-side errors are significant.

Finally, we incorporated IMU-derived gravity DV as an additional DV pair within the same DV-alignment framework, leading to extended Wahba and extended TLS formulations with an analytically determined gravity weight from accelerometer error statistics. Simulation results demonstrated that the gravity-DV augmentation further reduces overall attitude errors, particularly horizontal attitude errors (roll\&pitch), and improves solution availability under limited anchor availability, with the DOI Hessian matching weighting strategy providing additional yaw-error reduction when high-quality IMU data are available.

Overall, we provide a computationally efficient solution with covariance prediction ability for short-range AD using DoA measurements, augmented by an analytically weighted gravity-direction constraint. It further improves robustness and availability during static initialization, making it well suited for resource-constrained unmanned vehicles operating in complex environments.

\appendix[DoA Estimation Characteristics of a Three-Element Planar Array]\label{appendix}
This appendix presents a concrete implementation example of short-range array DoA estimation, which serves as the DoA front-end assumed in Sections~\ref{problem-formulation}--\ref{simulation}. Its role is twofold: 1) to illustrate how short-range arrays produce DVs with direction-dependent uncertainty, and 2) to provide a covariance model for the body-frame DV error $\boldsymbol{\Sigma}_{\boldsymbol{v}_k^b}$ used in the proposed Hessian-matching based weighting formulation \eqref{eq40} and in the simulation study.

\subsection{Array Geometry and PDoA Measurement Model}
\begin{figure}
\centering
\includegraphics[width=0.8\linewidth]{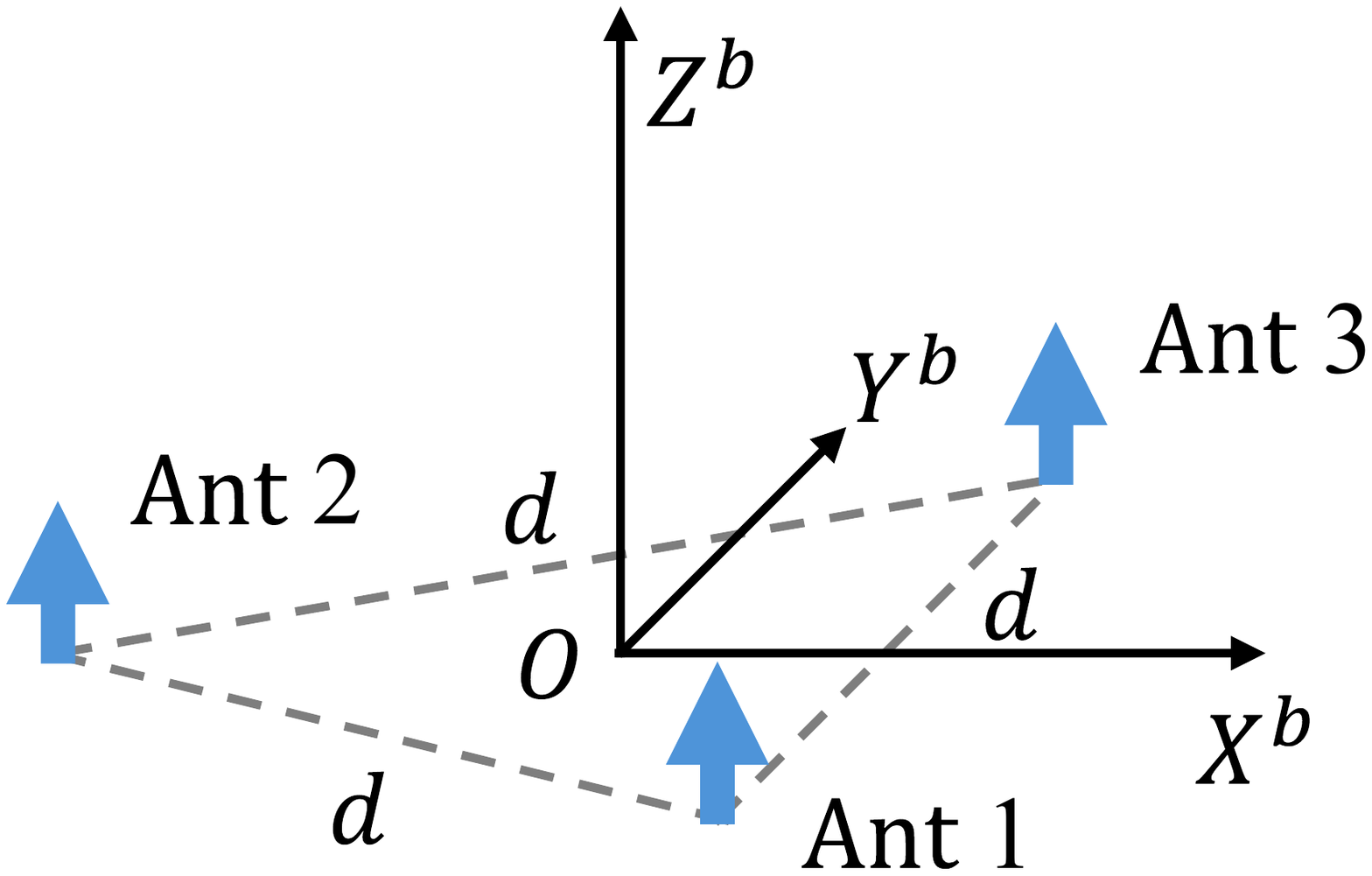}
\caption{ Geometry of the three-element planar antenna array with an equilateral-triangle configuration in the body frame.}
\label{fig10}
\end{figure}

As shown in Fig.~\ref{fig10}, we consider a compact three-element planar array arranged as an equilateral triangle in the body frame $\mathcal{B}$. The array center is located at the origin, and the antenna-element positions are
\begin{equation}
\begin{aligned}
\boldsymbol{p}^{b}_{\text{Ant}\ 1} &=
\left[\frac{d}{2\sqrt{3}},\ -\frac{d}{2},\ 0\right]^{\transpose}, \\
\boldsymbol{p}^{b}_{\text{Ant}\ 2} &=
\left[-\frac{d}{\sqrt{3}},\ 0,\ 0\right]^{\transpose}, \\
\boldsymbol{p}^{b}_{\text{Ant}\ 3} &=
\left[\frac{d}{2\sqrt{3}},\ \frac{d}{2},\ 0\right]^{\transpose},
\end{aligned}
\label{eq54}
\end{equation}
where $d$ denotes the inter-element spacing. This three-element configuration is attractive for lightweight unmanned vehicles due to its minimal hardware complexity: it provides two independent phase-difference measurements, which is the minimum required to estimate a 2D AoA under a far-field plane-wave approximation.

Radios enable accurate carrier PDoA extraction at the center frequency $f_c$ (with wavelength $\lambda$), which can be used for AoA estimation. Let $\boldsymbol{v}^{b}$ denote the incident unit DV in the body frame, parameterized by azimuth $\alpha$ and elevation $\beta$ as in \eqref{eq8}. The relative carrier phase observed at the $i$-th antenna with respect to the array center can be written as
\begin{equation}
\varphi_i = \frac{2\pi}{\lambda}\left(\boldsymbol{p}^{b}_{\text{Ant}\ i}\right)^{\transpose}\boldsymbol{v}^{b}.
\label{eq55}
\end{equation}
Taking antenna 3 as the reference, we form two independent PDoAs
\begin{equation}
\begin{aligned}
\Phi_1 &= \varphi_3-\varphi_1,\\
\Phi_2 &= \varphi_3-\varphi_2,
\end{aligned}
\label{eq56}
\end{equation}
which, after substituting \eqref{eq54} and \eqref{eq8} into \eqref{eq55}, yield the explicit mapping
\begin{equation}
\begin{aligned}
\Phi_1 &=
\frac{2\pi d}{\lambda}\cos\beta\,\sin\alpha,\\
\Phi_2 &=
\frac{2\pi d}{\lambda}\left(\frac{\sqrt{3}}{2}\cos\beta\,\cos\alpha+\frac{1}{2}\cos\beta\,\sin\alpha\right).
\end{aligned}
\label{eq57}
\end{equation}
In practice, $\boldsymbol{\Phi}=[\Phi_1,\Phi_2]^{\transpose}$ is corrupted by measurement noise and possible phase wrapping; the subsequent analysis focuses on the small-error regime (after unwrapping or within the unambiguous range).

\subsection{Closed-Form AoA Solution}
Given $\hat{\boldsymbol{\Phi}}$, closed-form estimates of azimuth and elevation can be obtained by inverting \eqref{eq57}:
\begin{equation}
\begin{aligned}
\alpha &= \text{atan2} \left(\sqrt{3}\Phi_1,\ 2\Phi_2-\Phi_1\right), \\
\beta  &= \arccos \left(\frac{\lambda}{\pi d}\sqrt{\frac{\Phi_1^2+\Phi_2^2-\Phi_1\Phi_2}{3}}\right).
\end{aligned}
\label{eq58}
\end{equation}
Linearizing \eqref{eq58} around the true $\boldsymbol{\Phi}$ gives the small perturbations
\begin{equation}
\begin{aligned}
\Delta \alpha &\approx \frac{\partial \alpha}{\partial \Phi_1}\Delta \Phi_1+\frac{\partial \alpha}{\partial \Phi_2}\Delta \Phi_2
= a_1\Delta \Phi_1+a_2\Delta \Phi_2,\\
\Delta \beta &\approx \frac{\partial \beta}{\partial \Phi_1}\Delta \Phi_1+\frac{\partial \beta}{\partial \Phi_2}\Delta \Phi_2
= b_1\Delta \Phi_1+b_2\Delta \Phi_2,
\end{aligned}
\label{eq59}
\end{equation}
where the coefficients are
\begin{equation}
\begin{aligned}
a_1 &= \frac{\sqrt{3}\Phi_2}{2(\Phi_1^2+\Phi_2^2-\Phi_1\Phi_2)}, \quad
a_2 = \frac{-\sqrt{3}\Phi_1}{2(\Phi_1^2+\Phi_2^2-\Phi_1\Phi_2)},\\
b_1 &= -\frac{1}{\sqrt{1-t^2}}\frac{\lambda}{2\pi d}
\frac{2\Phi_1-\Phi_2}{\sqrt{3(\Phi_1^2+\Phi_2^2-\Phi_1\Phi_2)}},\\
b_2 &= -\frac{1}{\sqrt{1-t^2}}\frac{\lambda}{2\pi d}
\frac{2\Phi_2-\Phi_1}{\sqrt{3(\Phi_1^2+\Phi_2^2-\Phi_1\Phi_2)}},\\
t &= \frac{\lambda}{\pi d}\sqrt{\frac{\Phi_1^2+\Phi_2^2-\Phi_1\Phi_2}{3}}.
\end{aligned}
\label{eq60}
\end{equation}

Let $\boldsymbol{\gamma}=[\alpha,\beta]^{\transpose}$ denote the 2D AoA vector. With the Jacobian
\begin{equation}
\boldsymbol{T}=
\begin{bmatrix}
a_1 & a_2\\
b_1 & b_2
\end{bmatrix},
\label{eq61}
\end{equation}
the AoA covariance follows from standard linear error propagation:
\begin{equation}
\boldsymbol{\Sigma}_{\boldsymbol{\gamma}}
= \boldsymbol{T}\,\boldsymbol{\Sigma}_{\boldsymbol{\Phi}}\,\boldsymbol{T}^{\transpose},
\label{eq62}
\end{equation}
where $\boldsymbol{\Sigma}_{\boldsymbol{\Phi}}$ is the PDoA covariance. In the simulations (Table~\ref{tab1}), we assume independent and identically distributed PDoA errors with variance $\sigma_{\Phi}^2$, i.e.,
\begin{equation}
\boldsymbol{\Sigma}_{\boldsymbol{\Phi}}
= \diag([\sigma_{\Phi}^2,\ \sigma_{\Phi}^2]).
\label{eq63}
\end{equation}

\subsection{Covariance analysis for DV Estimates}
The DoA estimates are converted to the body-frame unit DV via \eqref{eq8}. Let $\Delta\boldsymbol{\gamma}=[\Delta\alpha,\Delta\beta]^{\transpose}$. A first-order expansion gives
\begin{equation}
\Delta \boldsymbol{v}^{b} \approx \boldsymbol{S}\,\Delta \boldsymbol{\gamma},
\label{eq64}
\end{equation}
where the Jacobian $\boldsymbol{S}=\frac{\partial \boldsymbol{v}^{b}}{\partial \boldsymbol{\gamma}}$ is
\begin{equation}
\boldsymbol{S}=
\begin{bmatrix}
-\sin\alpha\cos\beta & -\cos\alpha\sin\beta\\
\cos\alpha\cos\beta  & -\sin\alpha\sin\beta\\
0                    & \cos\beta
\end{bmatrix}.
\label{eq65}
\end{equation}
Therefore, the body-frame DV covariance used in \eqref{eq40} is
\begin{equation}
\boldsymbol{\Sigma}_{\boldsymbol{v}^{b}}
= \boldsymbol{S}\,\boldsymbol{\Sigma}_{\boldsymbol{\gamma}}\,\boldsymbol{S}^{\transpose}.
\label{eq66}
\end{equation}

\begin{figure*}
\centering
\subfloat[\centering Azimuth $\alpha$]{\includegraphics[width=0.3\linewidth]{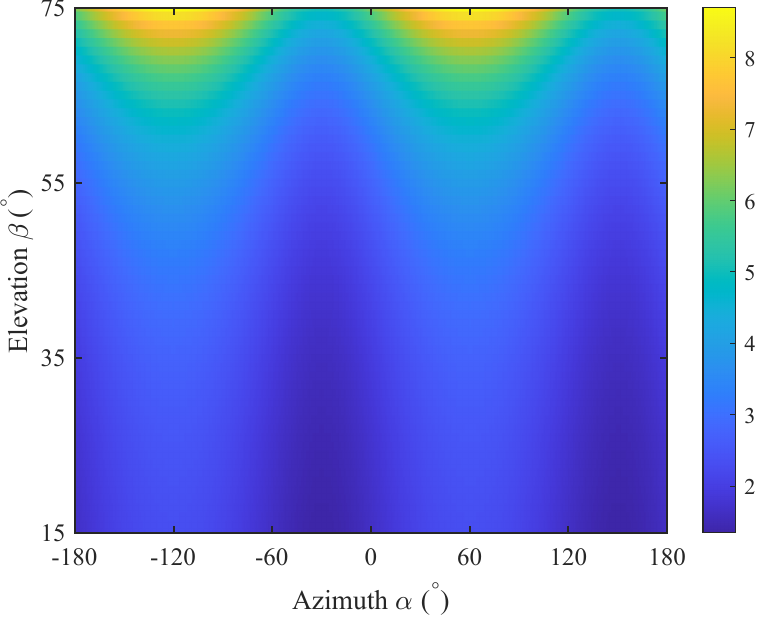}
\label{fig11_1}}
\hfil
\subfloat[\centering Elevation $\beta$]{\includegraphics[width=0.3\linewidth]{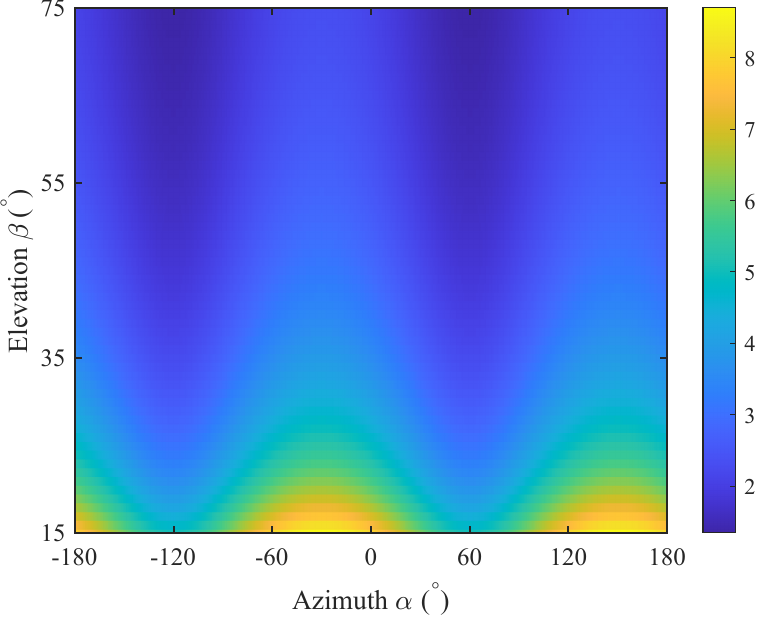}
\label{fig11_2}}
\hfil
\subfloat[\centering DV $\boldsymbol{v}^{b}$]{\includegraphics[width=0.3\linewidth]{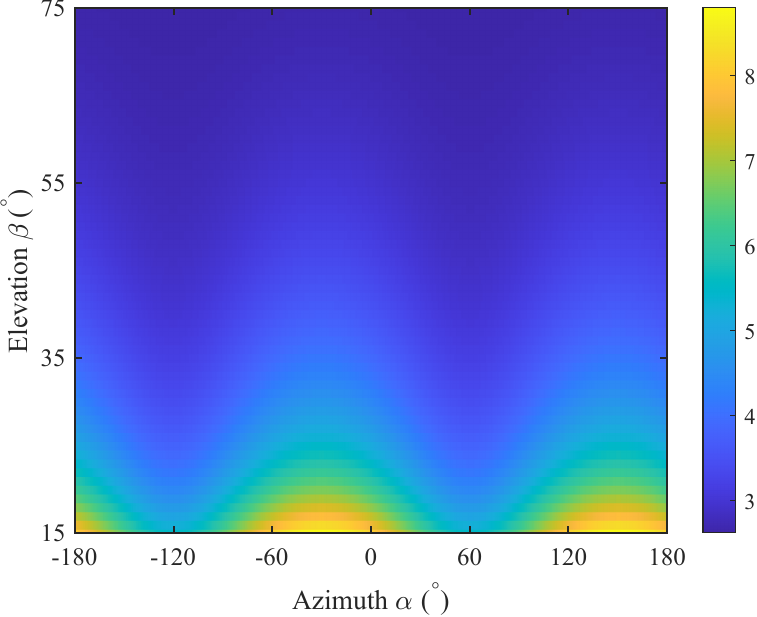}
\label{fig11_3}}
\caption{
Directional dependency of the theoretical AoA measurement errors and the corresponding DV estimation errors for the three-element planar antenna array. The results illustrate the anisotropic DoA accuracy induced by array geometry, which motivates the need for direction-dependent weighting in AD.}
\label{fig11}
\end{figure*}

As illustrated in Fig.~\ref{fig11}, both the theoretical AoA estimation errors and the induced DV estimation errors depend strongly on the incident direction, indicating non-uniform DoA accuracy across azimuth--elevation angles. This anisotropy must therefore be considered in AD, as also noted in previous studies \cite{wang2025attitude}. Equations \eqref{eq60}--\eqref{eq66} show that DV uncertainty generally depends on the incident direction through $(\alpha,\beta)$ and can be strongly anisotropic (see Fig.~\ref{fig11}). Moreover, due to the projection and unit-norm constraints, the resulting DV covariance can become ill-conditioned or even nearly singular for certain directions. These observations motivate the regularization in the TLS-consistent modeling adopted in the main text and justify the proposed Hessian-matching based weighting strategy.

\section*{Acknowledgments}
The authors acknowledge the use of the AI language model ChatGPT (OpenAI) to improve English language expression throughout the manuscript. No technical content or scientific ideas were generated by the AI.

\begin{IEEEbiography}[{\includegraphics[width=1in,height=1.25in,clip,keepaspectratio]{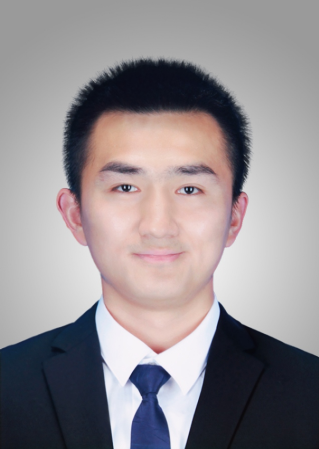}}]{Chenxin Tu} received the B.E. degree in electronic engineering from Tsinghua University, Beijing, China, in 2022, where he is currently pursuing the Ph.D. degree with the Department of Electronic Engineering.
His research interests include wireless localization and cooperative localization.
\end{IEEEbiography}

\begin{IEEEbiography}[{\includegraphics[width=1in,height=1.25in,clip,keepaspectratio]{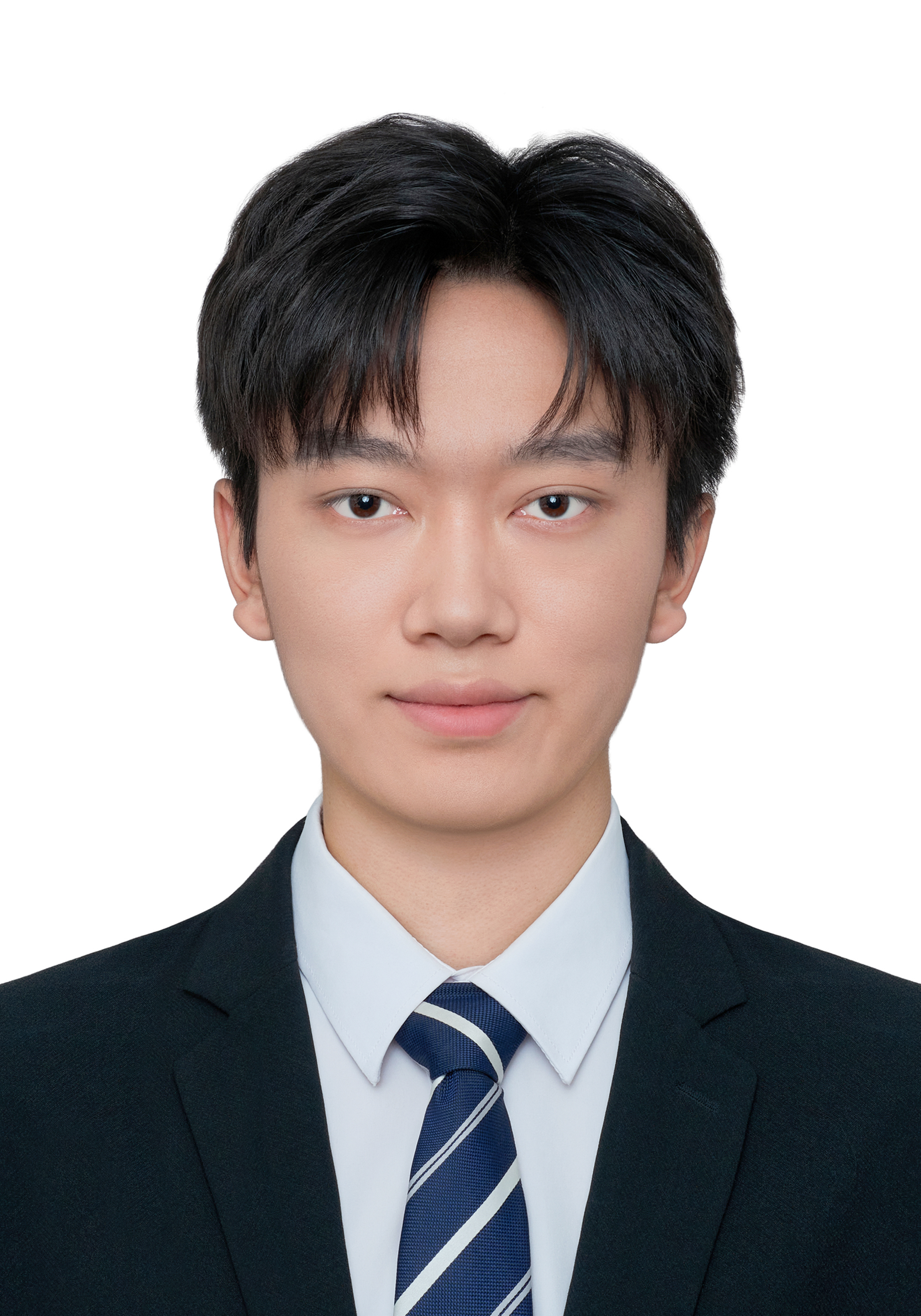}}]{Hengchuan Zhang} received the B.E. degree in electronic engineering from Tsinghua University, Beijing, China, in 2024, where he is currently pursuing the M.E. degree with the Department of Electronic Engineering. His research interests include wireless localization.
\end{IEEEbiography}

\begin{IEEEbiography}[{\includegraphics[width=1in,height=1.25in,clip,keepaspectratio]{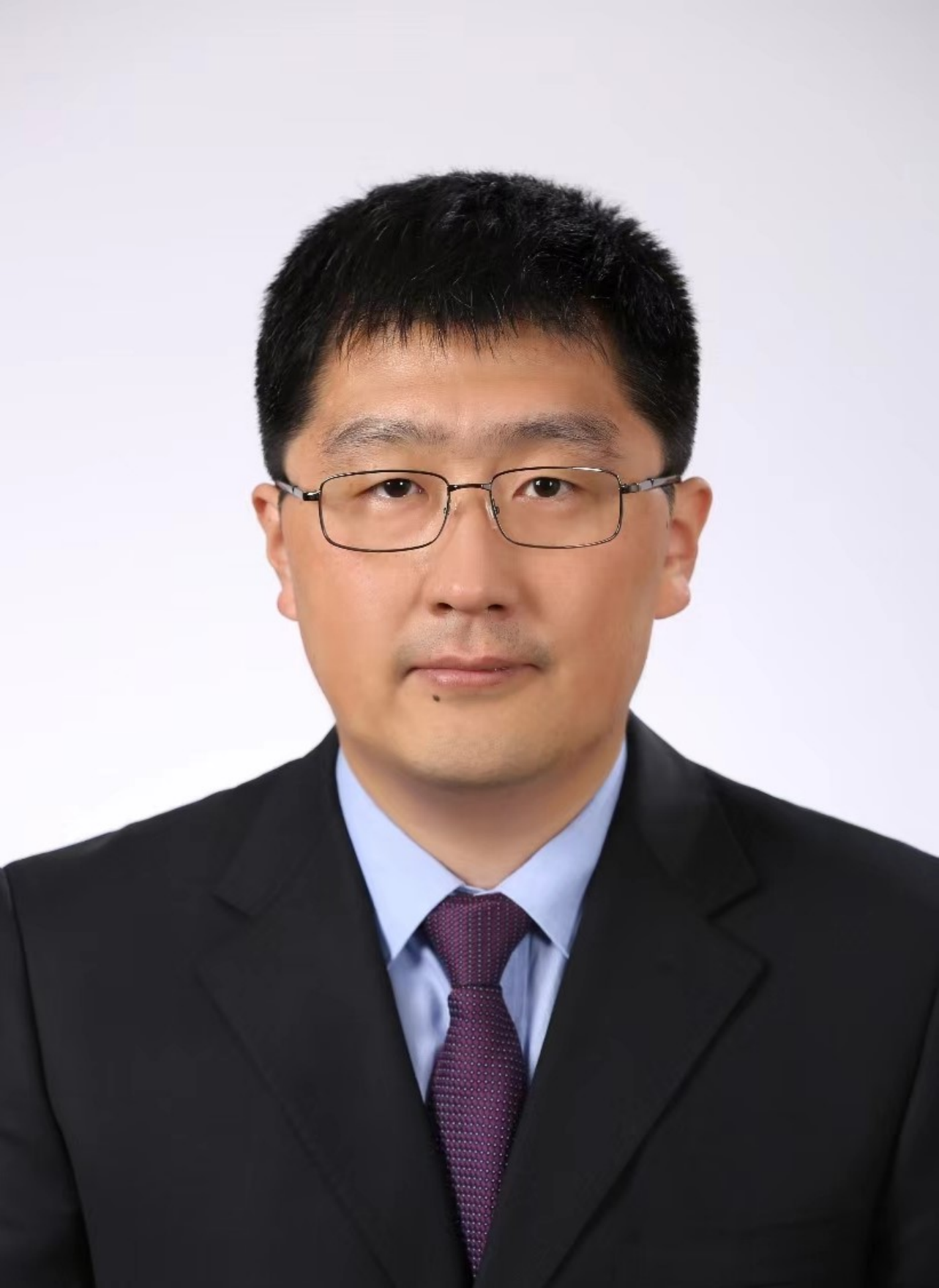}}]{Xiaowei Cui} received the B.S. and Ph.D. degrees in electronic engineering from Tsinghua University, Beijing, China, in 2000 and 2005, respectively. Since 2005, he has been with the Department of Electronic Engineering, Tsinghua University, where he is currently a Professor. 

His research interests include robust GNSS signal processing, multipath mitigation techniques, and high-precision positioning. He is a member of the Expert Group of China BeiDou Navigation Satellite System.
\end{IEEEbiography}

\begin{IEEEbiography}[{\includegraphics[width=1in,height=1.25in,clip,keepaspectratio]{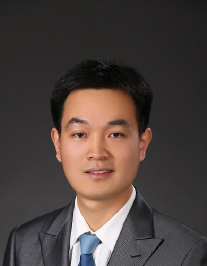}}]{Gang Liu} 
is an associate researcher at the Department of Electronic Engineering, Tsinghua University, China. His research interests include GNSS/INS integrated navigation techniques, and high-precision localization. He obtained both the BS and PhD degrees in instrument science and technology from Tsinghua University in 2007 and 2015, respectively.
\end{IEEEbiography}

\begin{IEEEbiography}[{\includegraphics[width=1in,height=1.25in,clip,keepaspectratio]{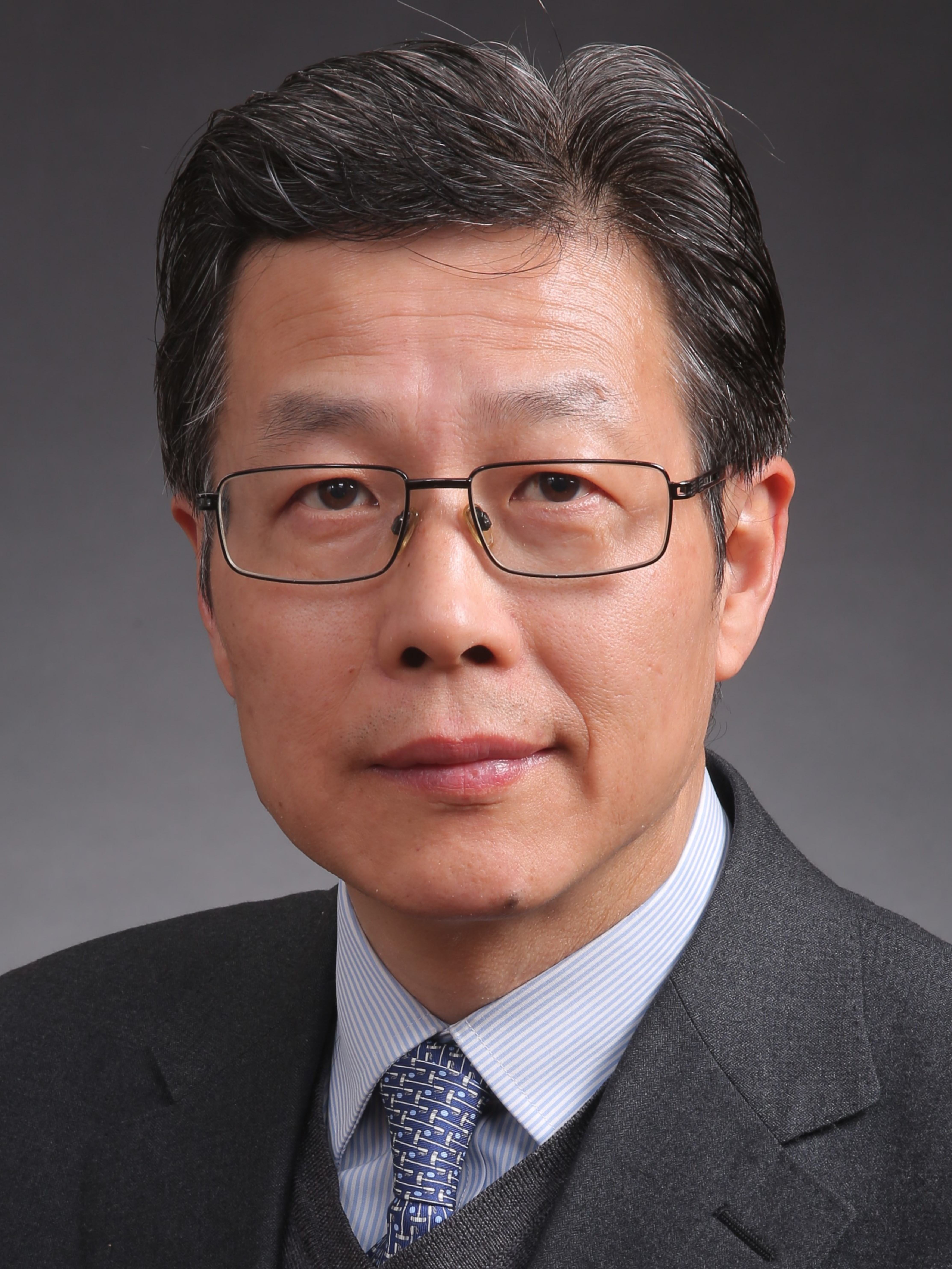}}]{Mingquan Lu} 
received the M.S. and Ph.D. degrees from University of Electronic Science and Technology of China, Chengdu, China. He is a Professor with the Department of Electronic Engineering, Tsinghua University, Beijing, China. He directs the PNT Research Center, which develops GNSS and alternative PNT technologies. He authored or co-authored 5 books and book chapters, published over 300 journal and conference papers, and hold nearly 100 patents. His current research focuses on GNSS signal processing, GNSS receiver development and emerging PNT technologies. 

Prof. Lu was a recipient of the ION Thurlow Award. He provided numerous services to the GNSS community, including an Associate Editor for Satellite Navigation and Journal of Navigation and Positioning. He is a Fellow of ION.
\end{IEEEbiography}


\begin{thebibliography}{45}

\bibitem{klemas2015coastal}V. V. Klemas,
``Coastal and environmental remote sensing from unmanned aerial vehicles: An overview,''
\emph{J. Coastal Res.}, vol. 31, no. 5, pp. 1260--1267, 2015.

\bibitem{li2022application}Y. Li, M. Liu, and D. Jiang,
``Application of unmanned aerial vehicles in logistics: a literature review,''
\emph{Sustainability}, vol. 14, no. 21, 2022, Art. no. 14473.

\bibitem{velusamy2021unmanned}P. Velusamy, S. Rajendran, R. K. Mahendran, S. Naseer, M. Shafiq, and J.-G. Choi,
``Unmanned aerial vehicles (UAV) in precision agriculture: Applications and challenges,''
\emph{Energies}, vol. 15, no. 1, 2021, Art. no. 217.

\bibitem{ellenberg2015use}A. Ellenberg, L. Branco, A. Krick, I. Bartoli, and A. Kontsos,
``Use of unmanned aerial vehicle for quantitative infrastructure evaluation,''
\emph{J. Infrastruct. Syst.}, vol. 21, no. 3, 2015, Art. no. 04014054.

\bibitem{oh2013formation}K.-K. Oh and H.-S. Ahn,
``Formation control and network localization via orientation alignment,''
\emph{IEEE Trans. Autom. Control}, vol. 59, no. 2, pp. 540--545, 2013.

\bibitem{senanayake2016search}M. Senanayake \emph{et al.},
``Search and tracking algorithms for swarms of robots: A survey,''
\emph{Robot. Auton. Syst.}, vol. 75, pp. 422--434, 2016.

\bibitem{hol2009tightly}J. D. Hol, F. Dijkstra, H. Luinge, and T. B. Schon,
``Tightly coupled UWB/IMU pose estimation,''
in \emph{Proc. IEEE Int. Conf. Ultra-Wideband}, Vancouver, BC, Canada, 2009, pp. 688--692.

\bibitem{feng2020kalman}D. Feng, C. Wang, C. He, Y. Zhuang, and X.-G. Xia,
``Kalman-filter-based integration of IMU and UWB for high-accuracy indoor positioning and navigation,''
\emph{IEEE Internet Things J.}, vol. 7, no. 4, pp. 3133--3146, 2020.

\bibitem{tu2025parameterized}C. Tu, X. Cui, G. Liu, S. Zhao, and M. Lu,
``Parameterized TDOA: TDOA estimation for mobile target localization in a time-division broadcast positioning system,''
\emph{IEEE Internet Things J.}, vol. 12, no. 13, pp. 24131--24147, 2025.

\bibitem{tu2025decoupled}C. Tu, X. Cui, G. Liu, and M. Lu,
``A decoupled localization and synchronization method for moving targets using sequential one-way TOA measurements,''
in \emph{Proc. IEEE Int. Conf. Commun. (ICC)}, Montreal, QC, Canada, 2025, pp. 2478--2484.

\bibitem{liu2022research}Y. Liu, R. Yu, Z. Xiong, and Y. Guo,
``Research on algorithms for multi-vector attitude determination,''
\emph{Math. Probl. Eng.}, vol. 2022, 2022, Art. no. 6137308, doi: \href{https://doi.org/10.1155/2022/6137308}{10.1155/2022/6137308}.

\bibitem{liebe1995star}C. C. Liebe,
``Star trackers for attitude determination,''
\emph{IEEE Aerosp. Electron. Syst. Mag.}, vol. 10, no. 6, pp. 10--16, Jun. 1995, doi: \href{https://doi.org/10.1109/62.387971}{10.1109/62.387971}.

\bibitem{lovera2006global}M. Lovera and A. Astolfi,
``Global magnetic attitude control of spacecraft in the presence of gravity gradient,''
\emph{IEEE Trans. Aerosp. Electron. Syst.}, vol. 42, no. 3, pp. 796--805, Jul. 2006, doi: \href{https://doi.org/10.1109/TAES.2006.248214}{10.1109/TAES.2006.248214}.

\bibitem{titterton2004strapdown}D. Titterton and J. L. Weston,
\emph{Strapdown Inertial Navigation Technology}, vol. 17.
London, U.K.: IET, 2004.

\bibitem{teunissen2017springer}P. J. G. Teunissen and O. Montenbruck, Eds.,
\emph{Springer Handbook of Global Navigation Satellite Systems}, vol. 10.
Cham, Switzerland: Springer, 2017.

\bibitem{meurer2012robust}M. Meurer, A. Konovaltsev, M. Cuntz, and C. H{\"a}ttich,
``Robust joint multi-antenna spoofing detection and attitude estimation using direction assisted multiple hypotheses RAIM,''
in \emph{Proc. ION GNSS}, Nashville, TN, USA, 2012, pp. 3007--3016.

\bibitem{daneshmand2014precise}S. Daneshmand, N. Sokhandan, and G. Lachapelle,
``Precise GNSS attitude determination based on antenna array processing,''
in \emph{Proc. ION GNSS+}, Tampa, FL, USA, 2014, pp. 2555--2562.

\bibitem{dominguez2016performance}E. Dom{\'i}nguez \emph{et al.},
``Performance evaluation of high sensitivity GNSS techniques in indoor, urban and space environments,''
in \emph{Proc. ION GNSS+}, Portland, OR, USA, 2016, pp. 373--393.

\bibitem{dotlic2017angle}I. Dotlic, A. Connell, H. Ma, J. Clancy, and M. McLaughlin,
``Angle of arrival estimation using decawave DW1000 integrated circuits,''
in \emph{Proc. 14th Workshop Posit. Navig. Commun. (WPNC)}, Bremen, Germany, 2017, pp. 1--6.

\bibitem{li2024high}Y. Li, H. Zhao, Y. Liu, T. Wang, J. Yu, and Y. Shen,
``High-accuracy 2-D AoA estimation using lightweight UWB arrays,''
in \emph{Proc. IEEE/RSJ Int. Conf. Intell. Robots Syst. (IROS)}, Abu Dhabi, UAE, 2024, pp. 3312--3317.

\bibitem{xiao2024research}K. Xiao, F. Hao, W. Zhang, N. Li, and Y. Wang,
``Research and implementation of indoor positioning algorithm based on bluetooth 5.1 AOA and AOD,''
\emph{Sensors}, vol. 24, no. 14, 2024, Art. no. 4579, doi: \href{https://doi.org/10.3390/s24144579}{10.3390/s24144579}.

\bibitem{wang2025attitude}C. Wang, X. Cui, G. Liu, and M. Lu,
``Attitude determination using GNSS DOA estimation under jamming,''
\emph{IEEE Trans. Aerosp. Electron. Syst.}, early access, 2025, doi: \href{https://doi.org/10.1109/TAES.2025.3630146}{10.1109/TAES.2025.3630146}.

\bibitem{chang2015total}G. Chang, ``Total least-squares formulation of Wahba's problem,'' \emph{Electron. Lett.}, vol. 51, no. 17, pp. 1334--1335, 2015.

\bibitem{wahba1965least}G. Wahba,
``A least squares estimate of satellite attitude,''
\emph{SIAM Rev.}, vol. 7, no. 3, pp. 409--409, 1965.

\bibitem{markley1988attitude}F. L. Markley,
``Attitude determination using vector observations and the singular value decomposition,''
\emph{J. Astronaut. Sci.}, vol. 38, pp. 245--258, 1988.

\bibitem{ahmed2017accurate}H. Ahmed and M. Tahir,
``Accurate attitude estimation of a moving land vehicle using low-cost MEMS IMU sensors,''
\emph{IEEE Trans. Intell. Transp. Syst.}, vol. 18, no. 7, pp. 1723--1739, Jul. 2017, doi: \href{https://doi.org/10.1109/TITS.2016.2627536}{10.1109/TITS.2016.2627536}.

\bibitem{decelis2018attitude}R. de Celis and L. Cadarso,
``Attitude determination algorithms through accelerometers, GNSS sensors, and gravity vector estimator,''
\emph{Int. J. Aerosp. Eng.}, vol. 2018, 2018, Art. no. 5394057, doi: \href{https://doi.org/10.1155/2018/5394057}{10.1155/2018/5394057}.

\bibitem{gebre2000gyro}D. Gebre-Egziabher, G. H. Elkaim, J. D. Powell, and B. W. Parkinson,
``A gyro-free quaternion-based attitude determination system suitable for implementation using low cost sensors,''
in \emph{Proc. IEEE Position Location Navig. Symp. (PLANS)}, San Diego, CA, USA, 2000, pp. 185--192, doi: \href{https://doi.org/10.1109/PLANS.2000.838301}{10.1109/PLANS.2000.838301}.

\bibitem{nocedal2006numerical}J. Nocedal and S. J. Wright,
\emph{Numerical Optimization}, 2nd ed.
New York, NY, USA: Springer, 2006.

\bibitem{niu2021wheel}X. Niu, Y. Wu, and J. Kuang,
``Wheel-INS: A wheel-mounted MEMS IMU-based dead reckoning system,''
\emph{IEEE Trans. Veh. Technol.}, vol. 70, no. 10, pp. 9814--9825, 2021.

\bibitem{IEEE802154z}
\emph{IEEE Standard for Low-Rate Wireless Networks--Amendment 1: Enhanced Ultra Wideband (UWB) Physical Layers (PHYs) and Associated Ranging Techniques}, IEEE Standard 802.15.4z-2020, 2020.

\bibitem{QorvoDW3000}\emph{DW3000: IEEE 802.15.4z UWB Transceiver Data Sheet}, Rev. 1.3,
Qorvo, Inc., Greensboro, NC, USA, 2020. [Online]. Available: \url{https://www.qorvo.com/products/p/DW3000}

\bibitem{analog2020adis16465}\emph{Precision MEMS IMU Module ADIS16465 Data Sheet}, Rev. C,
Analog Devices, Inc., Wilmington, MA, USA, 2020. [Online]. Available: \url{https://www.analog.com/media/en/technical-documentation/data-sheets/adis16465.pdf}

\bibitem{MPU6000Datasheet}
\emph{MPU-6000 and MPU-6050 Product Specification}, Rev. 3.4,
InvenSense Inc., Sunnyvale, CA, USA, Aug. 2013. [Online]. Available: \url{https://invensense.tdk.com/wp-content/uploads/2015/02/MPU-6000-Datasheet1.pdf}

\bibitem{zhao2025uwb}P. Zhao, H. Zhang, G. Liu, X. Cui, and M. Lu,
``A UWB-AOA/IMU integrated navigation system for 6-DoF indoor UAV localization,''
\emph{Drones}, vol. 9, no. 8, Aug. 2025, Art. no. 546.

\end{thebibliography}
\end{document}